\documentclass[conference]{IEEEtran}
\IEEEoverridecommandlockouts
\usepackage{cite}
\usepackage{amsmath,amssymb,amsfonts}
\newcommand{\R}{\mathbb{R}}
\usepackage{algorithmic}
\usepackage{graphicx,caption}
\usepackage{subfig}
\usepackage{textcomp}
\usepackage{xcolor}

\usepackage[hyphens]{url}
\usepackage{hyperref}
\usepackage[hyphenbreaks]{breakurl}

\usepackage[super]{nth}
\usepackage{listings}
\usepackage{dsfont}
\usepackage{multirow}
\usepackage[ruled,vlined]{algorithm2e}
\def\BibTeX{{\rm B\kern-.05em{\sc i\kern-.025em b}\kern-.08em
    T\kern-.1667em\lower.7ex\hbox{E}\kern-.125emX}}

\usepackage{tabularx}
\usepackage{tablefootnote}

\title{Accelerating Recommender Systems\\via Hardware ``scale-in''}

\author{\IEEEauthorblockN{Suresh Krishna$^{*}$\thanks{$^{*}$Equal contribution.}, Ravi Krishna$^{*}$}
	\IEEEauthorblockA{\textit{Electrical Engineering and Computer Sciences}\\
		\textit{University of California, Berkeley}\\
		{\tt\small \{suresh\_krishna, ravi.krishna\}@berkeley.edu}}
}

\begin{document}
\newcolumntype{L}[1]{>{\raggedright\arraybackslash}p{#1}}
\newcolumntype{C}[1]{>{\centering\arraybackslash}p{#1}}
\newcolumntype{R}[1]{>{\raggedleft\arraybackslash}p{#1}}

\maketitle
\thispagestyle{plain}
\pagestyle{plain}

\begin{abstract}
	In today's era of ``scale-out'', this paper makes the case that a specialized hardware architecture based
	on ``scale-in''{\textemdash}placing as many specialized processors as possible along with their memory systems and interconnect links within one or two boards in a rack{\textemdash}would offer the potential to boost large recommender system throughput
	by $\boldsymbol{12}$--$\boldsymbol{62\times}$ for inference and
	$\boldsymbol{12}$--$\boldsymbol{45\times}$ for training
	compared to the DGX-2 state-of-the-art AI platform,
	while minimizing the performance impact of distributing large models across multiple processors.
	By analyzing Facebook's representative model{\textemdash}Deep Learning Recommendation
	Model (DLRM){\textemdash}from a hardware architecture perspective,
	we quantify the impact on throughput of hardware parameters such as memory system design, collective communications
	latency and bandwidth, and interconnect topology.
	By focusing on conditions that stress hardware, our analysis reveals
	limitations of existing AI accelerators and hardware platforms.
\end{abstract}

\section{Introduction}
\label{intro}

Recommender Systems serve to personalize user experience in a variety of applications
including predicting click-through rates for ranking advertisements \cite{fbdlrm},
improving search results \cite{DeepRecSys}, suggesting friends and content on social
networks \cite{DeepRecSys}, suggesting food on Uber Eats \cite{ubereats},
helping users find houses on Zillow \cite{zillow}, helping contain information overload
by suggesting relevant news articles \cite{msmind},
helping users find videos to watch on YouTube \cite{youtubewatchnext} and movies
on Netflix \cite{netflix}, and several more real-world use cases \cite{usecases}.
An introduction to recommender system technology
can be found in \cite{googleclass}
and a set of best practices and examples for building recommender systems in \cite{msperso}.
The focus of this paper is recommendation systems that use neural networks,
referred to as Deep Learning RecSys,
or simply RecSys\footnote{While the term is often used to denote any recommender system,
it is specifically used for Deep Learning recommenders here.}.
These have been recently applied to a variety of areas with success \cite{fbdlrm}\cite{DeepRecSys}.

Due to their commercial importance{\textemdash}by improving the quality of Ads and content served to users,
RecSys directly drives revenues for hyperscalers, especially under cost-per-click billing{\textemdash}it is no surprise
that recommender systems consume the vast majority (${\sim}80\%$)
of AI inference cycles within Facebook's datacenters \cite{FBArch}; the situation
is similar \cite{DeepRecSys} at Google, Alibaba and Amazon.
In addition, RecSys training now consumes $>$50\% of AI training cycles within Facebook's
datacenters \cite{fbscaleupscaleout}.
Annual growth rates are $3.3\times$ for training \cite{fbscaleupscaleout} and
$2\times$ for inference \cite{fbdlinf}.
The unique characteristics of these workloads present challenges to datacenter
hardware including CPUs, GPUs and almost all AI accelerators.
Compared to other AI workloads such as computer vision or natural
language processing, RecSys tend to have
larger model sizes of up to 10TB \cite{aibox},
are memory access intensive \cite{DeepRecSys},
have lower compute burden \cite{fbdlinf}, and rely heavily
on CC\footnote{\textsc{Collective Communications.}} operations \cite{intelDLRM}.
These characteristics make RecSys a poor fit for many existing systems, as
described in Sec.~\ref{related}, and call for a new approach to accelerator HW architecture.

In this paper, various HW architectures are analyzed using Facebook's
DLRM \cite{DeepRecSys}
as a representative example and the resulting data are used to derive the characteristics
of \emph{RecSpeed}, an architecture optimized for running RecSys. Specifically, we:

\begin{itemize}
	\item Describe the DLRM workload in terms of its characteristics that impact
	HW throughput and latency.
	\item Identify HW characteristics such as memory system design, CC latency and bandwidth,
	and CC interconnect topology
	that are key determinants of upper bounds on RecSys throughput.
	\item Use a generalized roofline model that adds communication cost to memory and compute to
	show that specialized chip-level HW features can improve upper bounds on RecSys
	throughput $4{\times}$ by reducing latency and
	$7{\times}$ by improving bandwidth. It is known that a fixed-topology quadratic
	interconnect can offer CC all-to-all performance gains of $2.3{\times}$--$15{\times}$ \cite{fbscaleupscaleout}.
	\item Explain why AI accelerators for RecSys would benefit from
	supporting hybrid memory systems that combine
	multiple forms of DRAM.
	\item Evaluate the practical implementation of specialized HW
	for RecSys and show how this has the potential to improve throughput
	by $\boldsymbol{12}$--$\boldsymbol{62\times}$ for inference and
	$\boldsymbol{12}$--$\boldsymbol{45\times}$ for training
	compared to the NVIDIA DGX-2 AI platform.
\end{itemize}

\section{Related Work}
\label{related}

There are several published works that describe systems and chips for accelerating RecSys.
Compared to these, our work focuses on sweeping various hardware parameters for a
homogeneous\footnote{System where a single type of processor contributes the bulk of processing and communications capability.} system
in order to understand the impact of each upon
upper bound DLRM system throughput.
As such, we do not evaluate, from the standpoint of RecSys acceleration,
any of the other types of systems described below.

\subsection{Heterogeneous Platforms}
\label{related:heterogeneous}

\begin{figure*}[htbp]
	\captionsetup{width=.8\linewidth}
	\centerline{\includegraphics[clip, trim=0.5cm 0.5cm 1cm 0cm, width=2.0\columnwidth]{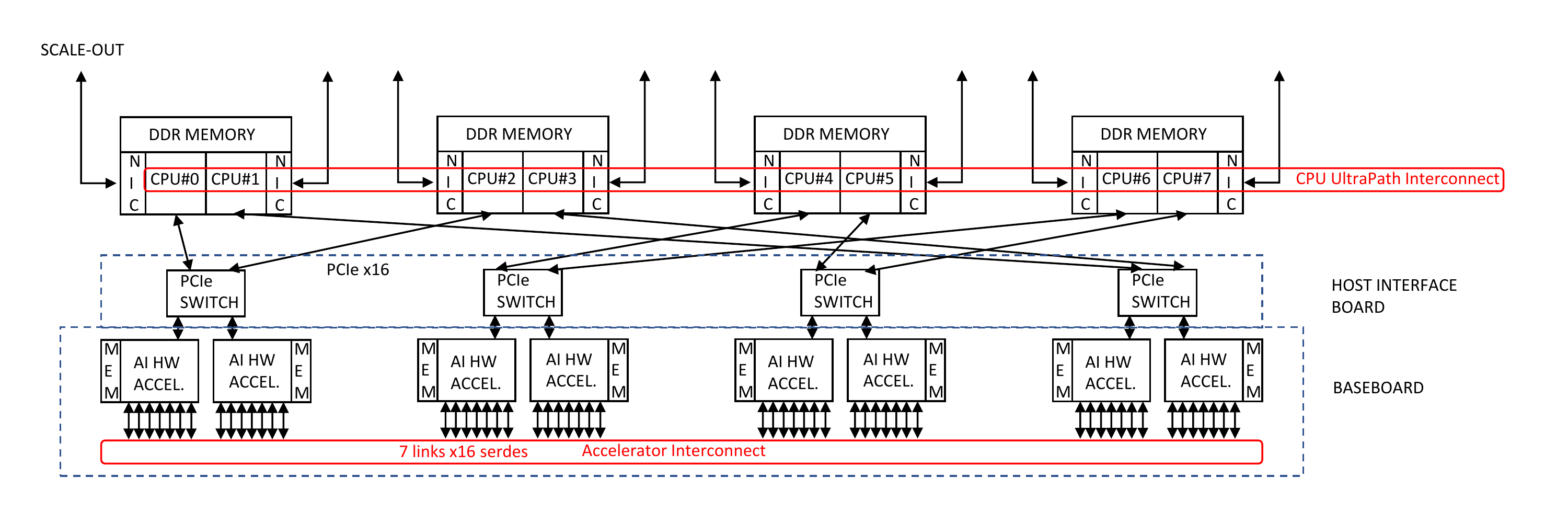}}
	\caption{FB Zion System: Major components and interconnect.}
	\label{fig:zion}
\end{figure*}

Facebook's Zion platform \cite{zionhc} is a specialized
heterogeneous\footnote{System that combines various processor types, each providing specialized compute and communications capability.} system
for training RecSys.
Fig.~\ref{fig:zion} shows the major components and interconnect of the Zion
platform, which are summarized in Table~\ref{table:fbzion}.
Zion offers the benefit of combining very large CPU memory to hold embedding tables
with high compute capability and fast interconnect from GPUs, along with
8 100GbE NICs for scale-out.
AIBox \cite{aibox} from Baidu is another heterogeneous system.
A key innovation of AIBox is the use of SSD memory to store the parameters of
a model up to 10TB in size; the system uses CPU memory as a cache for SSD.
The hit rate of this combination is reported to
increase as training proceeds, plateauing at about 85\% after 250 training
mini-batches. A single-node AIBox is reported to train a 10TB RecSys
with 80\% of the throughput of a 75-node MPI cluster at 10\% of the cost.

\begin{table}[htbp]
	\caption{Key Features of FB Zion \cite{zionhc}\cite{fbscaleupscaleout}.}
	
	\begin{center}
		\begin{tabular}{L{0.30\columnwidth} L{0.55\columnwidth}}
			\hline
			\textbf{Feature} & \textbf{Example of Implementation}\\
			\hline \hline
			CPU & 8x server-class processor such as Intel Xeon\\
			\hline
			CPU Memory speed & DDR4 DRAM, 6 channels/socket, up to 3200MHz, ${\sim}25.6$GB/s/channel\\
			\hline
			CPU memory capacity & Typical 1 DIMM/channel, up to 256GB/DIMM = 1.5TB/CPU\\
			\hline
			CPU Interconnect & UltraPath Interconnect, coherent\\
			\hline
			CPU I/O & PCIe Gen4 x16 per CPU, ${\sim}30$GB/s\\
			\hline
			AI HW Accelerator & Variable\\
			\hline
			Accelerator Interconnect & 7 links per accelerator, x16 serdes lanes/link, 25G, ${\sim}350$GB/s/card\\
			\hline
			Accelerator Memory & On-package, likely HBM\\
			\hline
			Accelerator Power & Up to 500W @ 54V/48V, support for liquid cooling\\
			\hline
			Scale-out & 1x NIC per CPU, typically 100Gb Ethernet\\
			\hline
		\end{tabular}
	\end{center}
	\label{table:fbzion}
\end{table}

\subsection{Homogeneous Platforms}
\label{related:homo}

\begin{table}[htbp]
	\caption{Key Features of NVIDIA DGX-A100 \cite{dgxa100specs}.}
	
	\begin{center}
		\begin{tabular}{L{0.25\columnwidth} L{0.6\columnwidth}}
			\hline
			\textbf{Feature} & \textbf{Example of Implementation}\\
			\hline \hline
			CPU & 2x AMD Rome 7742, 64 cores each\\
			\hline
			CPU Memory speed & DDR4 DRAM, 8 channels/socket, up to 3200MHz, ${\sim}25.6$GB/s/channel\\
			\hline
			CPU memory capacity & Typical 1 DIMM/channel, up to 256GB/DIMM = 2TB/CPU\\
			\hline
			AI HW Accelerator & 8x NVIDIA A100\\
			\hline
			Accelerator Interconnect & Switched all-to-all, NVLink3, 12 links/chip, ${\sim}300$GB/s bandwidth per chip\\
			\hline
			Accelerator Memory & HBM2 @ 2430MHz \cite{a100mem}, 40GB/chip, 320GB total\\
			\hline
			System Power & ${\sim}6.5$kW max.\\
			\hline
			Scale-out & 8x 200Gb/s HDR Infiniband\\
			\hline
		\end{tabular}
	\end{center}
	\label{table:dgxa100}
\end{table}

\begin{table}[htbp]
	\caption{Key Features of Intel/Habana HLS-1 \cite{habanahc}.}
	
	\begin{center}
		\begin{tabular}{L{0.25\columnwidth} L{0.6\columnwidth}}
			\hline
			\textbf{Feature} & \textbf{Example of Implementation}\\
			\hline \hline
			AI HW Accelerator & 8x Habana Gaudi\\
			\hline
			Accelerator Interconnect & Switched all-to-all, 10x 100Gb Ethernet per chip of which 7 are available for interconnect\\
			\hline
			Accelerator Memory & HBM, 32GB/chip, 256GB total\\
			\hline
			System Power & ${\sim}3$kW max.\\
			\hline
			Scale-out & 24x 100Gb Ethernet configured as 3 links per chip\\
			\hline
		\end{tabular}
	\end{center}
	\label{table:hls1}
\end{table}

NVIDIA's DGX-A100 and Intel/Habana's HLS-1 are two representative examples of homogeneous AI appliances.
Tables~\ref{table:dgxa100} and~\ref{table:hls1} respectively summarize their key characteristics.

\subsection{In/Near Memory Processing}
\label{related:nmp}

RecSys models tend to be limited by memory accesses to embedding table values that are then combined
using pooling operators\cite{recnmp}, which makes the integration of memory with embedding processing
an attractive solution.

The first approach for this is to modify a standard DDR4 DIMM by replacing its buffer chip
with a specialized processor that can handle embedding pooling operations.

\begin{figure}[htbp]
	\captionsetup{width=.8\linewidth}
	\centerline{\includegraphics[clip, trim=3cm 3.5cm 10cm 3cm, width=0.65\columnwidth]{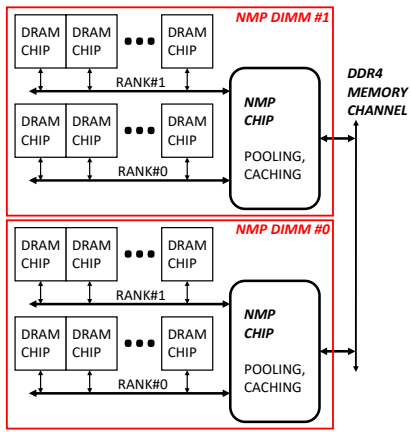}}
	\caption{Near-Memory Processing via Modified DIMM Buffer Chip. Example shows two DIMMs, each dual-rank, sharing a single
		memory channel.}
	\label{fig:nmp}
\end{figure}

Fig.~\ref{fig:nmp} shows the idea behind this concept for a scenario involving two
dual-rank DIMMs sharing one memory channel. In this example, the typical buffer chip found on
each DIMM is replaced by a specialized NMP (Near-Memory Processing) chip such as TensorDIMM\cite{tensordimm}
or RecNMP\cite{recnmp} that can access both ranks simultaneously, cache embeddings, and pool them prior
to transfer over the bandwidth-constrained channel.
For a high-performance server configuration with one dual-rank DIMM per memory channel,
simulations\cite{recnmp} indicate speedups
of $1.61\times$ to $1.96\times$.

It is also possible to provide a memory/compute module in a non-DIMM form factor.
An example is Myrtle.ai's SEAL Module \cite{myrtleseal}, an M.2 module specialized
for recommender systems embedding processing. This is built from Bittware's
250-M2D module \cite{bittware} that integrates a Xilinx
Kintex UltraScale+ KU3P FPGA along with two 32-bit channels of DDR4 memory.
Twelve such modules can fit within an OpenCompute Glacier Point V2 carrier board \cite{glacierv2},
in the space typically taken up by a Xeon CPU with its six 64-bit channels of DDR4 memory.

The second approach is to build a processor directly into a DRAM die.
UpMem \cite{upmem} has built eight 500MHz processors into an 8Gb DRAM.
Due to the lower speed of the DRAM process, the CPU uses a 14-stage pipeline
to boost its clock rate.
Several factors limit the performance gains available with this approach.
These include the limited number of layers of metal (e.g. 3) in DRAM semiconductor processes, the
need to place the processor far downstream from sensitive analog sense amplifier logic,
and the lag of DRAM processes behind logic processes.

\subsection{DRAM-less Accelerators}
\label{related:dramless}

DRAM-less AI accelerators include the Cerebras CS1 \cite{cs1}, Graphcore GC2 \cite{gc2citadel} and GC200 \cite{gc200}.
The CS1 and GC2 lack attached external DRAM, and the GC200 likely offers low-bandwidth access to such DRAM
since two DDR4 DIMMs are shared between four GC200s
via a gateway chip.

\subsection{Other approaches}
\label{related:others}

Centaur \cite{centaur} offloads embedding layer lookups and
dense compute to an FPGA that is co-packaged with a CPU via coherent links.
NVIDIA's Merlin \cite{nvmerlin}, while not a HW platform for RecSys, is an end-to-end solution
to support the development, training and deployment of RecSys on GPUs.
Intel \cite{intelDLRM} describes optimizations that improve DLRM throughput by $110{\times}$ on a single CPU socket,
to a level about $2{\times}$ that of a V100 GPU, with excellent scaling properties on clusters of up to 64 CPUs.

There is also a considerable body of work on domain-specific architectures for accelerating a broad set of AI applications
and there are several surveys of the field \cite{szetuto}\cite{mitsurvey}\cite{aihwelsevier}.
An up to date list of commercial chips can be found in \cite{aichips}.
Approaches using FPGAs are described in \cite{fowers}\cite{aifpga}, and \cite{tpuv3} describes a datacenter
AI accelerator.

\section{Overview of Recommender Systems}
\label{RecSysOverview}

In this section, we provide an overview of a representative RecSys, Facebook's DLRM \cite{fbdlrm},
from both application and algorithm/model perspectives,
including relevant deployment constraints and goals that will guide the development of our architecture.
An overview of several other RecSys can be found in \cite{DeepRecSys}.

\subsection{Black box model of RecSys; Inputs \& Output}
\label{RecSysOverview:Model}

The focus of this paper is a RecSys for rating an individual item of content, as opposed to RecSys that
process multiple items of content simultaneously \cite{ranknrate}.
Inputs to the RecSys are a description $u$ of a user and a description $c$ of a piece of content; the RecSys
outputs the estimated probability $P(u, c) \in (0,1)$ that the user will interact with the content in some specified way.
Both the user and the content are described by a set of \textbf{dense features} $\in\R^n$ and a set of \textbf{sparse features}
$\in \{0,1\}^m$.
Dense features are continuous, real-valued inputs such as the click-through rate of an Ad during the last week \cite{fb2014},
the average click-through rate of the user \cite{fb2014}, and the number of clicks already recorded for an Ad being ranked \cite{twitterads}.
Sparse (or categorical) features each correspond to a vector with a small number of 1-indices out of many 0s in multi-hot vectors, and represent information such as user ID, Ad category, and user history such as visited product pages or store pages \cite{alicat}\cite{alidin}.

From a conceptual standpoint, the RecSys could be run on every piece of content considered for a user and the resulting output
probabilities could then be used to determine which items of content to show that user to achieve business objectives such
as maximizing revenue or user engagement.
Note that the RecSys output may be combined with other components in order to decide what content to ultimately show the user,
as described in Sec.~\ref{RecSysOverview:Deployment}.

\subsection{Deployment Scenario}
\label{RecSysOverview:Deployment}

\begin{figure}[htbp]
	\captionsetup{width=.7\linewidth}
	\centerline{\includegraphics[clip, trim=2.5cm 1.5cm 17cm 0.5cm, width=0.9\columnwidth]{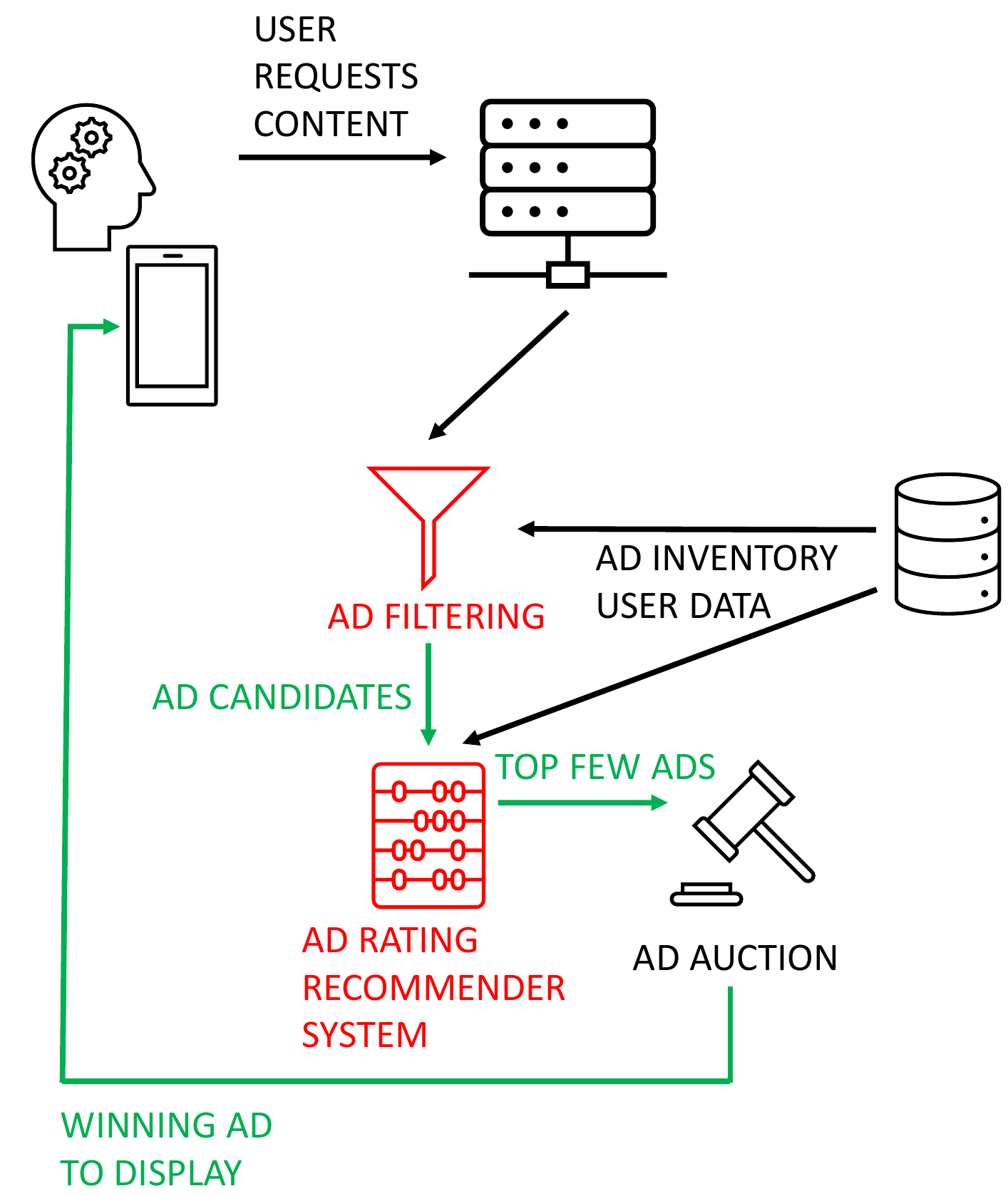}}
	\caption{Overview of multi-step Ad serving process consisting of filtering a large set of Ad content down to a small set of candidates, followed by ranking the candidates, the Ad auction, and displaying the winning Ad \cite{FBArch}\cite{DeepRecSys}\cite{varian}.}
	\label{fig:ad_process}
\end{figure}

A typical deployment scenario of a RecSys model
is illustrated in Fig.~\ref{fig:ad_process}:

\begin{enumerate}
	\item User loads a page, which triggers a request to generate personalized content. This request is sent from the user's device to the company's datacenter.
	\item Based on available content, an input \textbf{query} is generated, consisting of a set of $B$ (the \textbf{query size}) features, each one representing the piece of content and the user, and possibly their interactions. $B$ varies by recommendation model. There is typically a hierarchy of recommendation models whereby easier-to-evaluate models rank larger amounts of content first with high $B$, and then pass the top results to more complex models that rate smaller amounts of content with lower $B$ \cite{FBArch}. $B$ in the low to mid-hundreds is representative \cite{DeepRecSys}, with some $B$ as large as ${\sim}900$.
	\item This query is then passed on to the RecSys. This system outputs, for each of the $B$ pieces of potential content, the probability that the user will interact with that content in some way. In the case of advertising, this often means the probability that the user will click on the Ad. For video content, it could be metrics related to user engagement \cite{youtubewatchnext}.
	\item Based on these probabilities, the most relevant content is returned to the user, for instance ``the top tens of posts.'' \cite{FBArch}. However, for Ad ranking, the probabilities generated by the RecSys are first fed to an auction where they are combined with advertiser bids to select the Ad(s) that are ultimately shown to the user \cite{FBAdsAuctions} \cite{varian}.
\end{enumerate}

\subsection{System Constraints}
\label{RecSysOverview:Constraints}

RecSys operate under strict constraints. \\

\textbf{Inference Constraints}: The system must return the \textbf{most accurate results} within the (SLA and thus inviolate) \textbf{latency constraint} defined statistically in equation \ref{eqn:latency_constraint}, where $PPF(D_Q, P)$ is the percentage point function or inverse $CDF$, $D_Q$ is the distribution of the times to evaluate each query, $P$ is a percentile, such as \nth{99} or \nth{90}, and $C_{SLA}$ is the latency constraint, typically in the range of ``tens to hundreds of milliseconds'' \cite{FBArch}. Tail latencies are dependent on the QPS (queries per second) throughput 
of the serving system \cite{DeepRecSys};
one method to trade off QPS and tail latency which \cite{DeepRecSys} explores adjusting the query size.
\begin{equation}
	\label{eqn:latency_constraint}
	PPF(D_Q, P) \leq C_{SLA}
\end{equation}

\textbf{Training Constraints}: The system must train the \textbf{most accurate model} within the \textbf{minimum amount of time}. Recommendation models need to be retrained frequently in order to maintain accuracy when faced with changing conditions and user preferences \cite{fbhpca2018}, and time spent training a new model
is time when that model is not contributing to revenue.

Total deployment cost for the hardware needed to run the system is also a consideration.
This is measured as the \textbf{TCO}, or Total Cost of Ownership, of that hardware.

\subsection{Model Overview of DLRM}
\label{RecSysOverview:Models}

\begin{figure*}[htbp]
	\captionsetup{width=.8\linewidth}
	\centerline{\includegraphics[clip, trim=0 0 5cm 0.5cm, width=1.5\columnwidth]{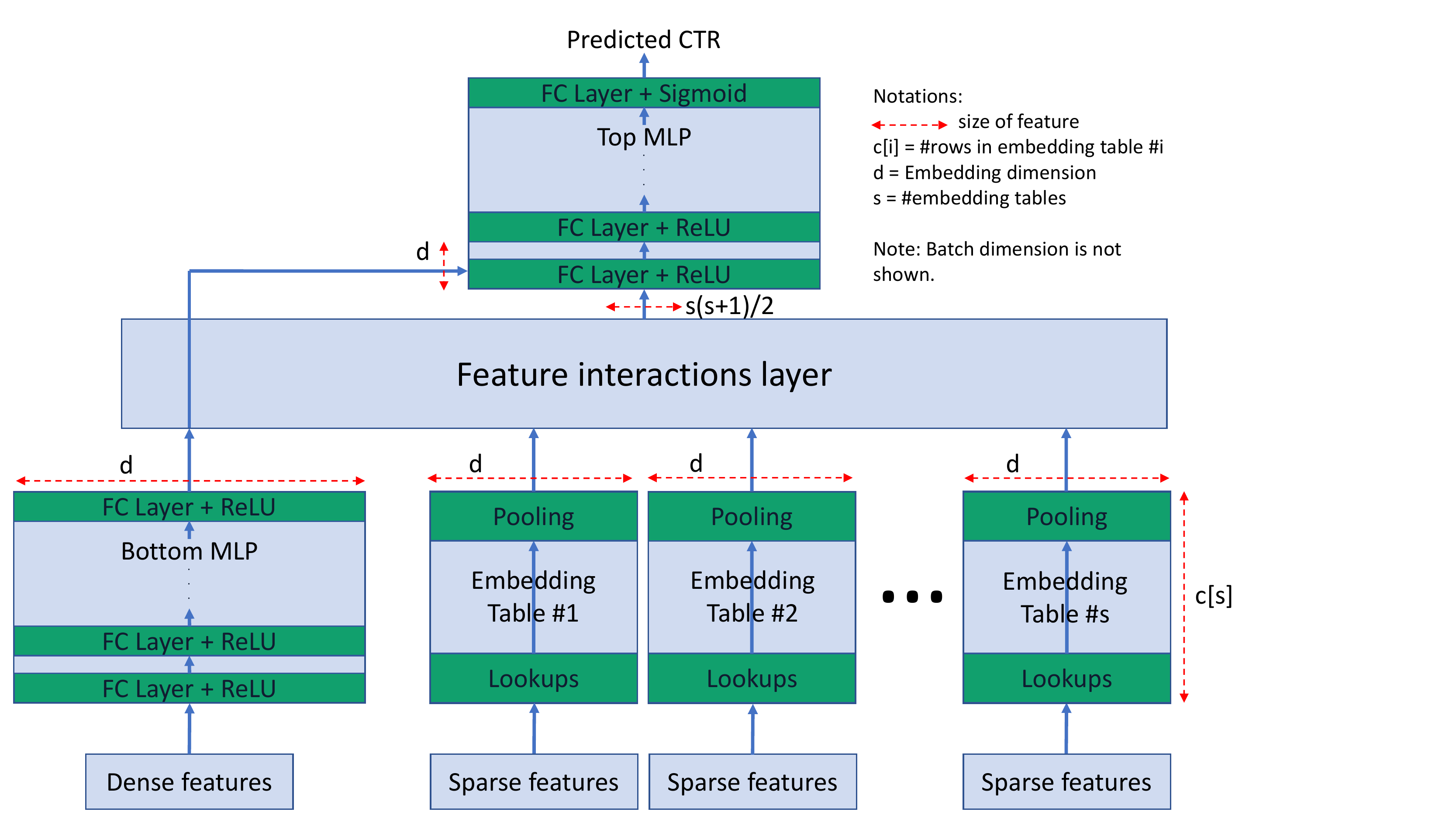}}
	\caption{DLRM Structure illustrating dense inputs and their associated bottom MLP, categorical inputs and embedding tables, feature interactions layer, and top MLP \cite{fbdlrm}.}
	\label{fig:dlrm}
\end{figure*}

The DLRM structure was open-sourced by Facebook in 2019 \cite{fbdlrm}. Fig.~\ref{fig:dlrm} illustrates the layers that comprise the DLRM model. DLRM is meant to be a reasonably general structure; for simplicity, the ``default'' implementation is used,
with sum pooling for embeddings, an FM\footnote{Factorization Machine.}-based \cite{rendlefm} feature interactions layer, and exclusion of the diagonal entries from the feature interactions layer output.

Inputs to DLRM are described in Sec.~\ref{RecSysOverview:Model}.
We note that each sparse feature is
effectively a set of indices into \textbf{embedding tables}. We now describe embedding tables and the other main components of DLRM.

\textbf{Embedding tables} are look-up tables, each viewed as a $c{\times}d$ matrix. Given a set of indices into a table, the corresponding rows are looked up, transposed to column vectors and combined into a single vector through an operation called \textbf{pooling}. Common pooling operators include sum and concatenation \cite{DeepRecSys}. An example of a sparse feature is user ID{\textemdash}a single index denoting the user for whom Ads are being ranked, corresponding to a single vector in an embedding table \cite{microsoftctr}.
We typically refer to the output of embedding tables as \textbf{embedding vectors},
and embedding tables themselves may be referred to as \textbf{embedding layers}.

\textbf{FC Layers}: The DLRM contains two Multi-Layer Perceptrons (MLPs), which consist of stacked (or composed) fully connected (FC) layers. The ``bottom MLP'' processes the dense feature inputs to the model, and the ``top MLP'' processes the concatenation of the feature interactions layer output and the bottom MLP output. DLRM uses ReLU activations for all FC layers except the last one in the top MLP which uses a sigmoid activation; this is to convert the 1-dimensional output of that layer to a click probability $\in (0, 1)$.

\textbf{Feature interactions layer}: This layer is designed to combine the information from the dense and sparse features.
The FM-based feature interactions layer forces the embedding dimension of every table and the output dimension of the final FC layer in the bottom MLP to all be equal.
After the dense features are first run through the bottom MLP, the output size is $b{\times}d$, where $b$ is the batch size being used for inference or training, and $d$ is the common embedding dimension. Further, each of the $s$ sparse feature embedding tables produces a $b{\times}d$ output after pooling. These are concatenated along a new dimension to construct a $b{\times}(s + 1){\times}d$ tensor which we will call $A$. Let $A' \in \mathcal{R}^{b{\times}d{\times}(s + 1)}$ be constructed by transposing the last two dimensions of $A$. $F$ is then calculated as the batch matrix multiplication of $A$ and $A'$ such that $F \in \mathcal{R}^{b{\times}(s + 1){\times}(s + 1)}$. Roughly half of these entries are duplicates, and those are discarded along with optionally the diagonal entries, which we opt to do. After flattening the two innermost dimensions, the result is the output matrix $F' \in \mathcal{R}^{b{\times}\frac{(s^2 + s)}{2}}$. This batch of vectors, after being concatenated with the bottom MLP output, is then fed to the top MLP. In algorithms \ref{alg:dlrm_fwd_steps} and \ref{alg:dlrm_bckwd_steps}, part of Sec.~\ref{DLRMHWArch:steps} where we cover the implementation of DLRM from a HW architecture perspective, for simplicity, this concatenation is considered as part of the feature interactions layer.

\subsection{RecSys vs. other AI models}
\label{RecSysOverview:arith}

The two key factors that set RecSys models apart from other AI models (such as those
for computer vision or natural language processing) are arithmetic intensity{\textemdash}the number of
compute operations relative to memory accesses{\textemdash}and model size. RecSys models are significantly larger
and have significantly lower arithmetic intensity as shown in Table~\ref{table:recsysai},
resulting in increased pressure on memory systems
and interconnect structure.

\begin{table*}[htbp]
	\caption{Recommender System vs. Other AI Models. Data from Table 1 of \cite{fbdlinf}.}
	
	\begin{center}
		\begin{tabular}{llccc}
			\hline
			\textbf{Category} & \textbf{Model Type} & \textbf{Size \#Parms} & \textbf{Arithmetic Intensity} & \textbf{Max. Live Activations}\\
			\hline \hline
			Computer Vision & ResNeXt101-32x4-48 & 43-829M & Avg. 188-380 & 2.4-29M\\
			\hline
			Language & Seq2Seq GRU/LSTM & 100M-1B & 2-20 & $>$100K\\
			\hline
			Recommender & Fully Connected Layers & 1-10M & 2-200 & $>$10K\\
			\hline
			\textbf{Recommender} & \textbf{Embedding Layers} & $\boldsymbol{>}$\textbf{10B} & \textbf{1-2} & $>$10K\\
			\hline
		\end{tabular}
	\end{center}
	\label{table:recsysai}
\end{table*}

\section{DLRM from a Hardware Architecture perspective}
\label{DLRMHWArch}

\subsection{Distributed model setup}
\label{DLRMHWArch:modelsplit}

The characteristics of recommender systems require a combination of model parallelism
and data parallelism.
On a large homogeneous system, dense parameters such as FC layer weights are \textbf{copied} onto every processor. However, embedding tables are \textbf{distributed} across the memory of all processors,
such that no parameters are replicated. This arises from the fact that embedding tables can easily reach from
several 100 GBs to multiple TBs in size \cite{paddlebox}.

While each processor can compute the dense part of the model for a batch, it cannot look up the embeddings specified by the sparse features, because some of those embeddings are stored in the memory of other processors. Thus, each processor needs to query other processors for embedding entries. This leads to the communication patterns outlined in Sec.~\ref{DLRMHWArch:communication}.

There are multiple ways to split up the embedding tables.
The distribution and (if beneficial) replication of tables across processors must be optimized
to avoid system bottlenecks by
evening out memory access loading across the memory systems of the various processors.
One extreme is ``full sharding'' of the tables, where tables are split up at a vector-level
across the attached memory systems of multiple processors to get very close to uniformly distributing table lookups{\textemdash}at the cost of increased communication.
This increased communication and the resulting stress on the HW is due to the fact that, with full sharding, each processor is prevented from doing local pooling of embeddings, thus requiring unpooled vectors to be communicated, of which there are many more than pooled vectors,
as described in Sec.~\ref{RecSysOverview:Models}.
As such, many systems today attempt to fit entire tables into the memory of single processors (which we call ``no sharding''),
or to break up the tables as little as possible as shown on slide 6 of \cite{zionhc}.

\subsection{CC operations}
\label{DLRMHWArch:communication}

The following CC operations are key to distributed RecSys throughput.
For more information on CC operations, please see \cite{ccbook}.
In all of the scenarios described below, there are $n$ processors, numbered $1, 2, \cdots, n$.

\textbf{All-to-all}: This CC primitive essentially implements a ``transpose'' operation. Each processor starts out with with $n$ pieces of data. Supposing $A_{ij}$ denotes the $j$th piece of data currently residing on the $i$th processor, then the all-to-all operation will move $A_{ij}$ such that it is instead on the $j$th processor and is ordered as the $i$th piece of data there. This operation is useful when each processor needs to send some data to every other processor.

\textbf{All-reduce}: All-reduce is an operation which replaces local data on each processor with the sum of data across all processors. If processor $i$ contains data $A_i$, then after the all-reduce operation, all processors will contain $A_R = \sum_{i = 1}^{n} A_i$. Efficient all-reduce algorithms exist for ring interconnects \cite{ccbook} which are popular in AI systems; however, all-reduce can also be implemented in equivalent time by first performing a reduce-scatter operation and then an all-gather operation.

\textbf{Reduce-scatter}: This operation may be best described as an all-to-all operation followed by a ``local reduction.'' Similar to the all-to-all setup, the starting point is $A_{ij}$ as the $j$th piece of data on processor $i$. Where all-to-all will result in this being the $i$th piece of data on processor $j$, reduce-scatter performs an extra reduction, such that the only piece of data on processor $j$ at the end is $\sum_{i = 1}^{n} A_{ij}$.

\textbf{All-gather}: In this operation, the starting point is a single piece of data $A_i$ on each of the $n$ processors, and the result of the operation is to have every $A_i$ on every processor. For example, with 4 processors initially containing $A_1$, $A_2$, $A_3$, and $A_4$, respectively, after the all-gather operation, each processor will contain all of $A_1, A_2, A_3,$ and $A_4$.

\subsection{Detailed steps for DLRM Inference and Training}
\label{DLRMHWArch:steps}

Please refer to the Appendix for a description of operators and an in-depth view of the steps involved in DLRM inference an training.

\subsection{Key HW performance factors}
\label{DLRMHWArch:keyhwperf}

Our performance model (Sec.~\ref{perf:perfmodel}) helps identify several HW characteristics that
are major determinants of throughput for the RecSys that we evaluate.
As could be expected from the DLRM operations described in the previous three sections,
these are CC performance, memory system performance for embedding table lookups,
compute performance for running the dense portions of the model,
and on-chip buffering. The following sections examine each of these factors in more detail.

\subsubsection{CC Performance}
\label{DLRMHWArch:keyhwperf:cc}

RecSys make extensive use of CC for exchanging embedding
table indices, embedding values, and for gradient averaging during training backprop.
Table~\ref{table:hwcc} summarizes HW factors that impact RecSys CC
performance.
In particular, all-to-all with its smaller message sizes is especially sensitive to latency 
and scales poorly with interconnect such as rings \cite{fbscaleupscaleout}.
Various lower bounds on the latency and throughput of these operators have been derived
for representative system architectures \cite{ccbook}.

\begin{table}[htbp]
	\caption{Summary of HW Collective Communications factors for RecSys.}
	
	\begin{center}
		\begin{tabular}{L{0.30\columnwidth} L{0.55\columnwidth}}
			\hline
			\textbf{Factor} & \textbf{Impact}\\
			\hline \hline
			Per-processor bandwidth & Sets upper-bounds on CC all-reduce and all-to-all throughput\\
			\hline
			Interconnect topology & Major determinant of all-to-all throughput\\
			\hline
			Ring interconnect & Well-suited for all-reduce, poorly-suited for all-to-all\\
			\hline
			CC latency & Particularly important for all-to-all due to smaller message sizes\\
			\hline
		\end{tabular}
	\end{center}
	\label{table:hwcc}
\end{table}

For the purposes of HW analysis on homogeneous\footnote{Systems where all processors share equally in the communications and arithmetic components of CC.} systems, we note the following \cite{ccbook}:

\begin{itemize}
	\item For an all-to-all with data volume $V$ and $n$ processors,
	the lower bound on
	the amount of data sent and received by each processor is $V\times\frac{(n-1)}{n}$.
	\item Similarly for an all-reduce,
	the lower bound is
	$2{\times}V\times\frac{(n-1)}{n}$.
	\item The above bounds impose a minimum
	per-processor data volume that must be transferred. \emph{As a consequence, the bandwidth per processor
	will limit overall all-to-all and all-reduce throughput, even as more processors are added to a system.}
	As a rule of thumb, for a large system with a per-processor interconnect bandwidth $BW$,
	the maximum achievable system-wide all-to-all and all-reduce throughputs are roughly $BW$ and $\frac{BW}{2}$.
\end{itemize}

The above rule of thumb works well for NVIDIA's DGX-2 system built from 16 V100 processors \cite{nvswitch}
since it achieves an ``all-reduce bandwidth\footnote{Data transfer rate in each direction, per processor, during all-reduce.}'' of {$\sim$}118GB/s with a maximum per-processor bandwidth (Fig. ~9 of \cite{nvswitch})
of 150GB/s from six NVLink2 interconnects per V100 chip, for an efficiency of 79\%
of the theoretical peak.

\begin{figure}[htbp]
	\captionsetup{width=.8\linewidth}
	\centerline{\includegraphics[clip, trim=2cm 5cm 2cm 5.5cm, width=\columnwidth]{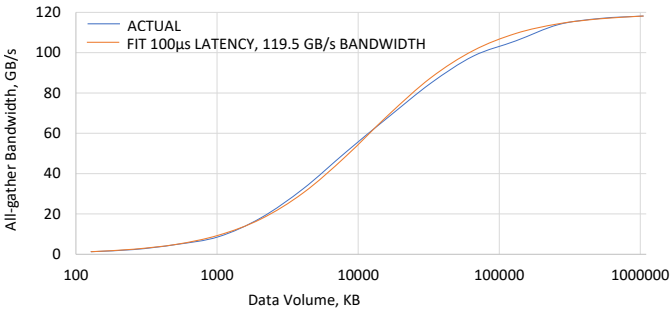}}
	\caption{DGX-2 All-gather bandwidth, measured values \cite{dgxlat} compared to simple latency/bandwidth model. All-gather time for smaller message sizes is latency dominated.}
	\label{fig:dgx2agbw}
\end{figure}

Fig.~\ref{fig:dgx2agbw} illustrates the impact
of latency on CC throughput. Due to the ${\sim}$100$\mu$s latency
for all-gather on the DGX-2,
the time to perform an all-gather for smaller message
sizes is latency dominated.
Latencies for a few current systems are shown in Table~\ref{table:latencies}.

\begin{table}[htbp]
	\caption{Latencies of interest.}
	
	\begin{center}
		\begin{tabular}{L{0.6\columnwidth} L{0.25\columnwidth}}
			\hline
			\textbf{System} & \textbf{Latency incl. SW overhead}\\
			\hline \hline
			NVIDIA DGX-2 CC All-reduce & Est. ${\sim}50\mu$s \cite{dgxlat}\\
			\hline
			NVIDIA DGX-2 CC All-gather & Est. ${\sim}100\mu$s \cite{dgxlat}\\
			\hline
			NVIDIA DGX-2 NVLink2 point to point & ${\sim}10\mu$s \cite{dgxlat}\\
			\hline
			NVIDIA DGX-2 PCIe point to point & ${\sim}30\mu$s \cite{dgxlat}\\
			\hline
			Graphcore GC2, 16-IPU system, single-destination Gather & ${\sim}25\mu$s \cite{gc2citadel}\\
			\hline
			Graphcore GC2, 16-IPU system, single-destination Reduce & ${\sim}14.5\mu$s \cite{gc2citadel}\\
			\hline
		\end{tabular}
	\end{center}
	\label{table:latencies}
\end{table}

The other important factor for CC, available per-chip peak bandwidth, is summarized in Table~\ref{table:hwbw}.
Note that Broadcom's Tomahawk 4 switch chip supports an order of magnitude more bandwidth than
any AI HW chip, \emph{demonstrating the headroom available to improve per-chip bandwidth for AI applications}.

\begin{table}[htbp]
	\caption{HW communication peak bandwidth for various chips.}
	
	\begin{center}
		\begin{tabular}{L{0.4\columnwidth} L{0.45\columnwidth}}
			\hline
			\textbf{Chip} & \textbf{Peak HW bandwidth, each direction, per chip}\\
			\hline \hline
			CPU: Intel Xeon Platinum 8180 & 62GB/s UPI, 48GB/s PCIe aggregate \cite{intelDLRM}, \cite{intel8180}, \cite{intelxsp}\\
			\hline
			GPU: NVIDIA A100 & NVLink3, 300GB/s \cite{a100spec}\\
			\hline
			FPGA: Achronix Speedster AC7t1500 & 400GB/s via 112G SerDes \cite{achrods}\\
			\hline
			AI HW: Graphcore GC200 & 320GB/s via 10x IPU-Links \cite{gc200server}\\
			\hline
			AI HW: Intel/Habana Gaudi & 125GB/s via 10x 100GbE \cite{habanahc}\\
			\hline \hline
			\textbf{Network Switch: Broadcom Tomahawk 4} & \textbf{3,200GB/s} \cite{th4}\\
			\hline
		\end{tabular}
	\end{center}
	\label{table:hwbw}
\end{table}

\subsubsection{Memory System Performance}
\label{DLRMHWArch:keyhwperf:mem}

This section refers to external DRAM memory attached to an AI HW accelerator or CPU.
RecSys applies considerable pressure on the memory system due to the large number of
accesses to embedding tables. Furthermore, these accesses have limited spatial locality (but some
temporal locality) \cite{recnmp}, resulting in scattered memory accessed of 64B-256B in size \cite{FBArch} that exhibit poor DRAM page hit characteristics. Multiple ranks per DIMM and internal banks and bank
groups per memory die help by increasing parallelism, within memory device timing parameters
such as command issue rates and on-die power distribution limitations.

\begin{table}[htbp]
	\caption{External Memory systems for select AI HW.}
	
	\begin{center}
		\begin{tabular}{L{0.35\columnwidth} L{0.50\columnwidth}}
			\hline
			\textbf{Chip} & \textbf{Memory System}\\
			\hline \hline
			Intel Xeon CPU & DDR4, 6 channels for Xeon Gold, up to 1.5TB \cite{samsung256gb}\\
			\hline
			NVIDIA A100 & HBM, 5 stacks, 40GB total\\
			\hline
			NVIDIA TU102 RTX2080Ti & GDDR6, 11 chips, 11GB total\\
			\hline
			Habana Goya & DDR4, 2 channels, 16GB\\
			\hline
			Graphcore GC2 & None, 300MB on-chip\\
			\hline
			Graphcore GC200 & 900MB on-chip, 2x DDR4 channels for 4 chips \cite{gc200}\\
			\hline
			Cerebras CS1 & None, all memory is on-wafer, 18GB total\\
			\hline
		\end{tabular}
	\end{center}
	\label{table:hwmem}
\end{table}

\begin{figure}[htbp]
	\captionsetup{width=.8\linewidth}
	\centerline{\includegraphics[clip, trim=1.5cm 4.5cm 1.5cm 4.5cm, width=0.9\columnwidth]{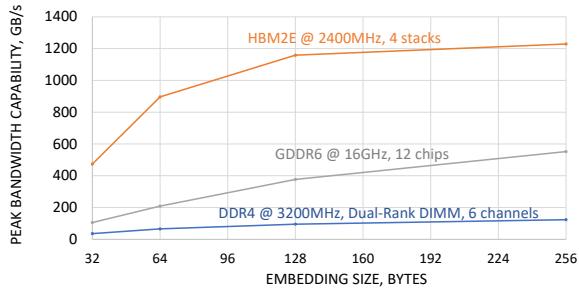}}
	\caption{Peak Random Embedding Access Bandwidth for common memory systems based on memory timing parameters,
		assuming auto-precharge. Data transfer frequency shown; device clock is half for DDR4 \& HBM, one-eighth for GDDR6. DDR4 memory systems have far lower performance than HBM for embedding lookups.}
	\label{fig:drambw}
\end{figure}

\begin{figure}[htbp]
	\captionsetup{width=.8\linewidth}
	\centerline{\includegraphics[clip, trim=1.5cm 3cm 1.5cm 3cm, width=0.8\columnwidth]{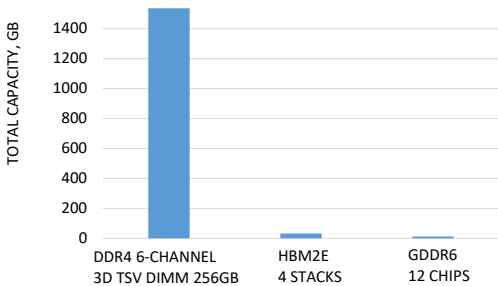}}
	\caption{Total capacity of common memory systems.}
	\label{fig:memcap}
\end{figure}

Table~\ref{table:hwmem} shows memory types for a few representative AI HW chips
and Fig.~\ref{fig:drambw} show achievable effective memory bandwidth
for various memory system configurations and embedding sizes.
HBM has considerably higher performance for random embedding accesses,
while the typical 6-channel DDR4 server CPU memory system has far lower performance,
especially for smaller embedding sizes. However HBM and GDDR6
suffer from limited capacity compared to DDR4 as shown in Fig.~\ref{fig:memcap}.

\subsubsection{Compute Performance}
\label{DLRMHWArch:keyhwperf:flops}

\begin{table}[htbp]
	\caption{Peak Compute capability of various chips.}
	
	\begin{center}
		\begin{tabular}{L{0.45\columnwidth} L{0.40\columnwidth}}
			\hline
			\textbf{Chip} & \textbf{FLOPS capability}\\
			\hline \hline
			CPU: Intel Xeon Platinum 8180 & 4.1 TFLOPS FP32 \cite{intelDLRM}\\
			\hline
			GPU: NVIDIA A100 & 19.5 TFLOPS FP32, 312 TFLOPS FP16/32 \cite{a100spec}\\
			\hline
			FPGA: Achronix Speedster AC7t1500 & 3.84 TFLOPS FP24 @ 750MHz \cite{achromlp}, \cite{achrods}\\
			\hline
		\end{tabular}
	\end{center}
	\label{table:flopshw}
\end{table}

Table~\ref{table:flopshw} shows available compute capability of various chips.
For the specific workload that we analyze, compute capability is not a limiting factor
(see Sec.~\ref{perf:models}) and this is believed to be the case for many
RecSys workloads when run on specialized AI accelerators \cite{DeepRecSys}.

\subsubsection{On-Chip Buffering}
\label{DLRMHWArch:keyhwperf:buf}

\begin{table}[htbp]
	\caption{Uses for on-chip buffer memory.}
	
	\begin{center}
		\begin{tabular}{L{0.35\columnwidth} L{0.50\columnwidth}}
			\hline
			\textbf{Item} & \textbf{Buffering memory}\\
			\hline \hline
			Dense weights for data parallelism & Model-dependent, replicated across each chip\\
			\hline
			Embeddings for one mini-batch & Dependent on embedding table sizes, mini-batch size, and (temporal) locality across mini-batch\\
			\hline
			Working buffers for data transfers & Used to overlap processing and transfers for input features and embedding lookups\\
			\hline
			Data during training & Activations, gradients, optimizer state \cite{zero}\cite{googopt}\\
			\hline
		\end{tabular}
	\end{center}
	\label{table:buf}
\end{table}

Buffering memory serves several purposes as shown in
Table~\ref{table:buf}. While some of these are straightforward to estimate{\textemdash}such as the number of
weights in the dense layers of a model{\textemdash}others are harder to quantify,
such as the number of unique embedding values across one or multiple mini-batches.
Analyses by Facebook \cite{recnmp} indicate hit rates of 40\% to 60\% with a 64MB cache.

\subsection{Improving existing AI HW accelerators for RecSys}
\label{DLRMHWArch:currenthw}

\begin{table}[htbp]
	\caption{Improving existing AI HW accelerators for RecSys.}
	
	\begin{center}
		\begin{tabular}{L{0.25\columnwidth} L{0.6\columnwidth}}
			\hline
			\textbf{Chip} & \textbf{Potential Changes}\\
			\hline \hline
			NVIDIA A100 & Increase on-chip memory for buffering; Add DDR memory support; Enable sixth HBM stack; 
			support deeper stacks; reduce compute capability to fit die area budget\\
			\hline
			Graphcore GC200 & Add external high-speed DRAM support\\
			\hline
			Cerebras CS1 & Change from mesh to fully connected topology, Add external high-speed DRAM support\\
			\hline \hline
			All of the above & Increase I/O bandwidth; Add HW support for CC; Add dual external DRAM support (Sec.~\ref{recspeed:design})\\
			\hline
		\end{tabular}
	\end{center}
	\label{table:changes}
\end{table}

Table~\ref{table:changes} illustrates potential changes to enhance the performance of existing
AI HW accelerators on RecSys workloads.

\section{Performance Model}
\label{perf}

\subsection{Representative DLRM Models}
\label{perf:models}

Our choice of representative model is Facebook's DLRM-RM2 \cite{DeepRecSys}.
In order to reveal limitations of various HW platforms, two model configurations are analyzed.
These are positioned at the low and high end of batch size and embedding entry size; specifically, 200 and 600 as the batch size points\footnote{Roughly 200 is the median query size; 600 is significantly farther out in the query size distribution \cite{DeepRecSys}.} and 64B and 256B as the embedding size points\footnote{Facebook has noted that embedding sizes in bytes are typically 64B-256B \cite{recnmp}.}.
We refer to the 200 batch size/64B embedding size combination as  \textbf{``Small batch/Small embeddings''},
while the 600 batch size/256B combination is \textbf{``Large batch/Large embeddings''}.

Similarly, the two extremes of table distribution are analyzed. ``Unsharded'' refers to each table being able to fit within
the memory attached to an AI accelerator, such that only pooled embeddings need be exchanged.
``Sharded'' refers to ``full sharding'' (see section \ref{DLRMHWArch:modelsplit}) where each table is fully distributed across the attached memory of every accelerator{\textemdash}a
worst-case scenario. The reality will likely fall between these two extremes.

Note that the same batch sizes are used for both our inference and training performance models and that all parameters are stored in FP16 for both inference and training. As the size of the tables for the
production model is not publicly available, we assume that
the model is large enough to occupy the memory of all chips
in the system. This is a reasonable assumption based on other
recommendation systems \cite{aibox}.

\begin{table}[htbp]
	\caption{DLRM-RM2 configurations.}
	
	\begin{center}
		\begin{tabular}{L{0.40\columnwidth} L{0.45\columnwidth}}
			\hline
			\textbf{Parameter} & \textbf{Value(s)}\\
			\hline \hline
			Number of embedding tables & 40\\
			\hline
			Lookups per table & 80\\
			\hline
			Embedding size & 32 FP16 = 64B (small), 128 FP16 = 256B (large)\\
			\hline
			Number of dense features & 256\\
			\hline
			Bottom MLP & 256-128-32-Embedding dimension\\
			\hline
			Top MLP & Interactions-512-128-1\\
			\hline
			Feature Interactions & Dot products, exclude diagonal\\
			\hline
			FLOPS/Inference & ${\sim}$1.40 MFLOPs (small), ${\sim}$2 MFLOPs (large)\\
			\hline
			Batch size & 200 (small), 600 (large)\\
			\hline
		\end{tabular}
	\end{center}
	\label{table:modelfeat}
\end{table}

\subsection{Overview of Performance Model}
\label{perf:perfmodel}

We have developed a performance model that computes the time, memory usage and communications overhead
for each of the steps detailed in Sec.~\ref{DLRMHWArch:steps}.
Our model takes as input a specific DLRM configuration as described in Sec.\ref{perf:models}
as well as various parameters that describe batch size, embedding table sharding,
processing engine capabilities and configuration,
memory system configuration, memory device timing parameters from
vendor datasheets, communication network latencies and bandwidths, and
system parameters that control the level of overlap and concurrency within the HW.
In order to maximally stress the HW, our model assumes zero (temporal) locality within the embedding access stream.

Results are reported in Sec.~\ref{eval} for the stated configurations,
assuming the HW and system exploit maximal overlap within a batch for inference
by grouping memory accesses and overlapping memory activity with communications where possible,
and for training by pipelining the collective communications with backpropagation computations
and parameter updates. For embedding parameter updates during training, the originally looked up embeddings are buffered on-chip, thereby only requiring a write to update them instead of a read-modify-write. In particular, sufficient on-chip buffering to support
the above capabilities is assumed{\textemdash}\emph{doing so over-estimates the performance
capabilities of most current AI accelerator HW.}

\section{Evaluation of System Performance}
\label{eval}

\subsection{Systems Parameters}
\label{eval:homoparms}

We focus our performance evaluation on homogeneous systems, where one type of
processor provides the bulk of compute and communications capability.

\begin{table}[htbp]
	\caption{Ranges of Key Parameters for Homogeneous Systems.}
	
	\begin{center}
		\begin{tabular}{L{0.25\columnwidth} L{0.6\columnwidth}}
			\hline
			\textbf{Parameter} & \textbf{Range}\\
			\hline \hline
			CC all-to-all & Latency of $0.5\mu$s to $10\mu$s, Bandwidth 100 to 1000GB/s\\
			\hline
			CC all-reduce & Latency of $0.5\mu$s to $10\mu$s, Bandwidth 100 to 1000GB/s\\
			\hline
			Bandwidth per chip & 100 to 1000GB/s aggregated across all links\\
			\hline
			\#chips/system & 8\\
			\hline
			Compute capability & 200 TFLOPS FP16\\
			\hline
			Memory system & 6 stacks of HBM2E per chip @ 2400 MHz\\
			\hline
		\end{tabular}
	\end{center}
	\label{table:homofeat}
\end{table}

Table~\ref{table:homofeat} shows the range of parameters that we consider for our reference homogeneous system.
In terms of CC performance, these ranges are, for the most part, significantly in excess of what is supported by state of the art AI HW accelerators.
This is consistent with our goal of showing the benefit of further optimizing these parameters.
In particular, the CC latency range is lower than measured numbers
for NVIDIA's DGX-2 system (Sec.~\ref{recspeed:perf}).
The range of per-chip bandwidth spans popular training accelerators as well as
about $3\times$ beyond NVLink3.

\subsection{Inference}
\label{eval:homoinf}

\begin{figure*}%
	\centering
	\subfloat[Small batch, Small embeddings, Unsharded]{{\includegraphics[clip, trim=3cm 0cm 0cm 0cm, width=0.8\columnwidth]{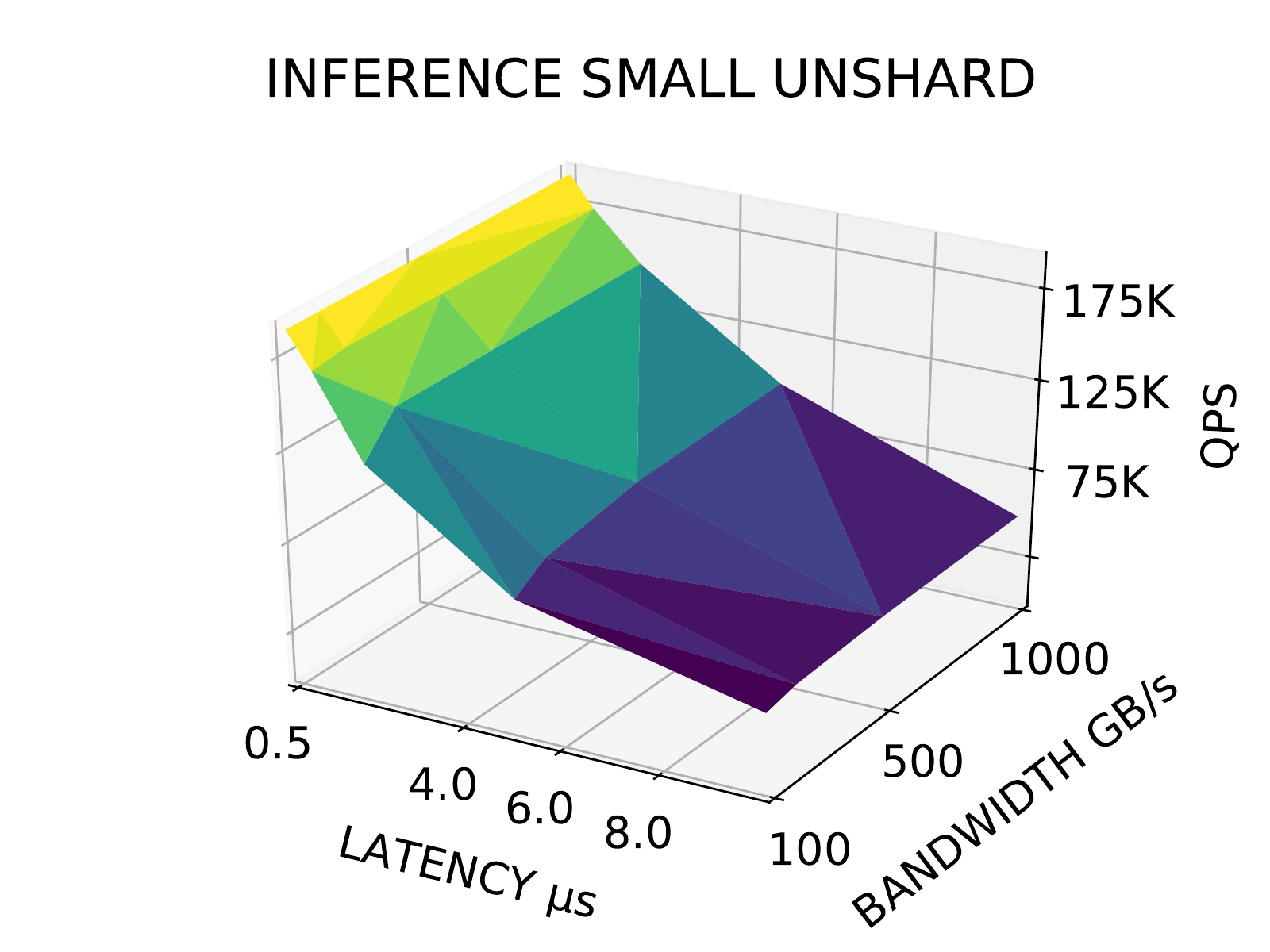} }}%
	\qquad
	\subfloat[Small batch, Small embeddings, Sharded]{{\includegraphics[clip, trim=3cm 0cm 0cm 0cm, width=0.8\columnwidth]{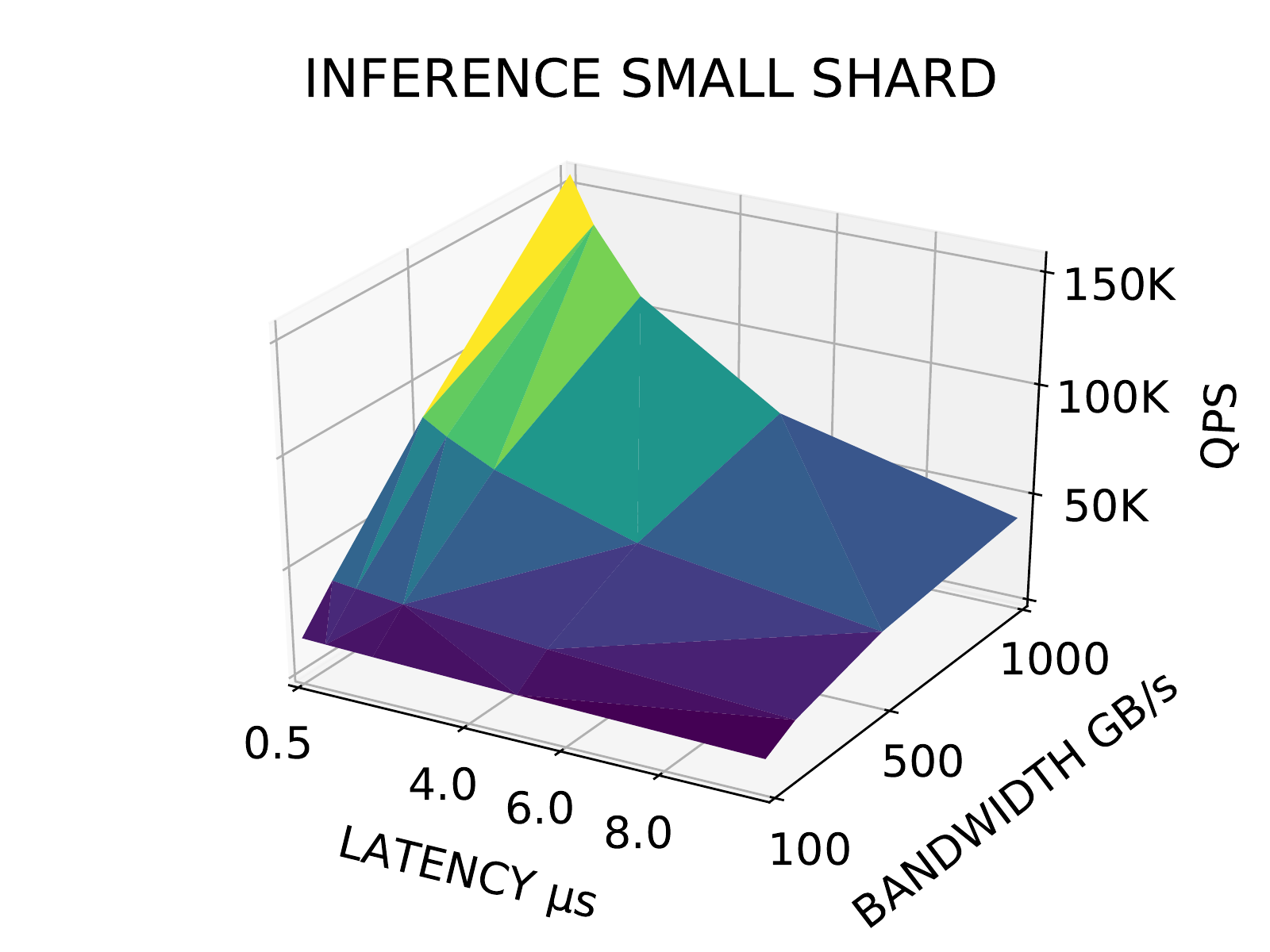} }}%
	\newline
	\subfloat[Large batch, Large embeddings, Unsharded]{{\includegraphics[clip, trim=3cm 0cm 0cm 0cm, width=0.8\columnwidth]{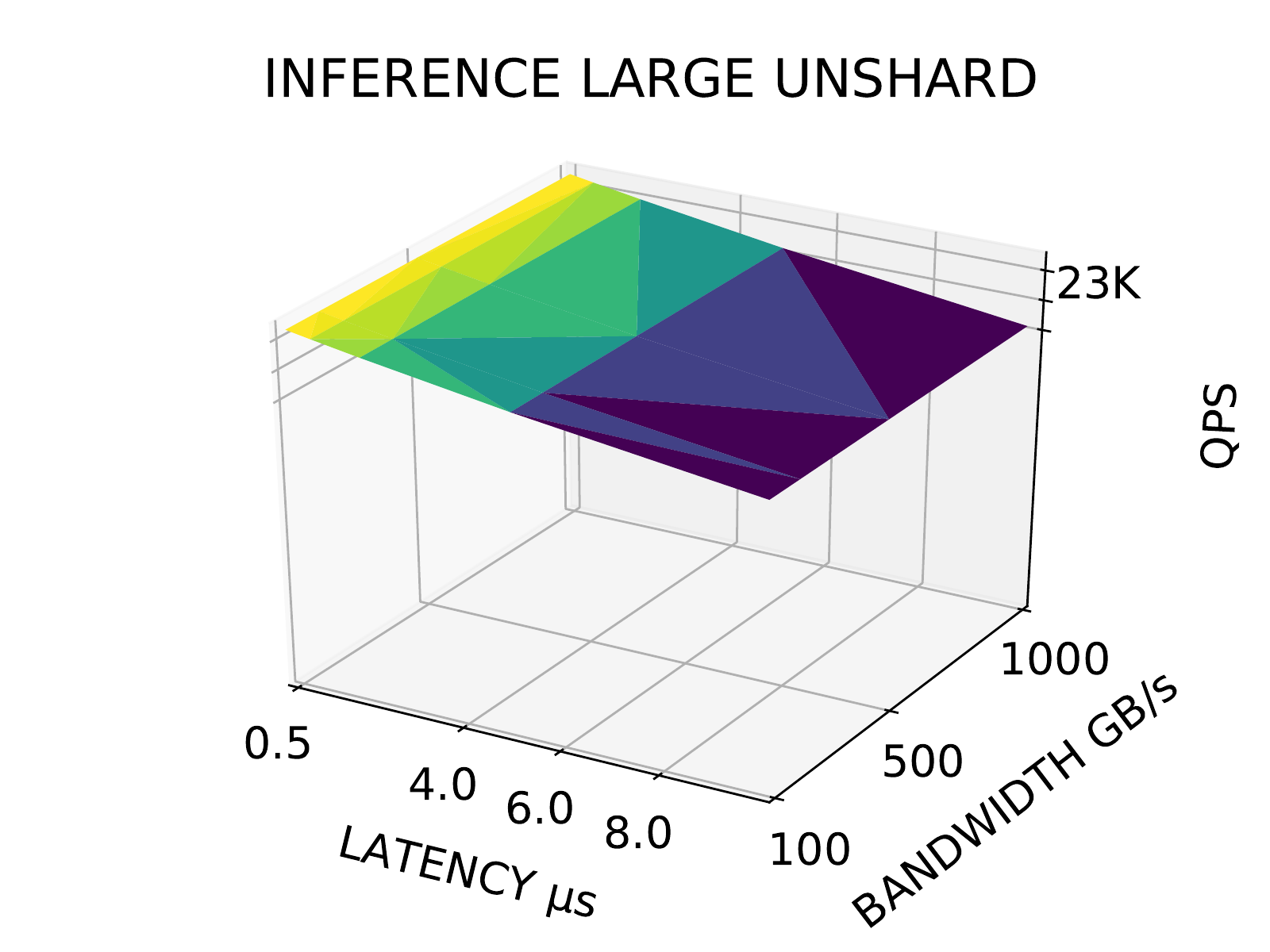} }}%
	\qquad
	\subfloat[Large batch, Large embeddings, Sharded]{{\includegraphics[clip, trim=3cm 0cm 0cm 0cm, width=0.8\columnwidth]{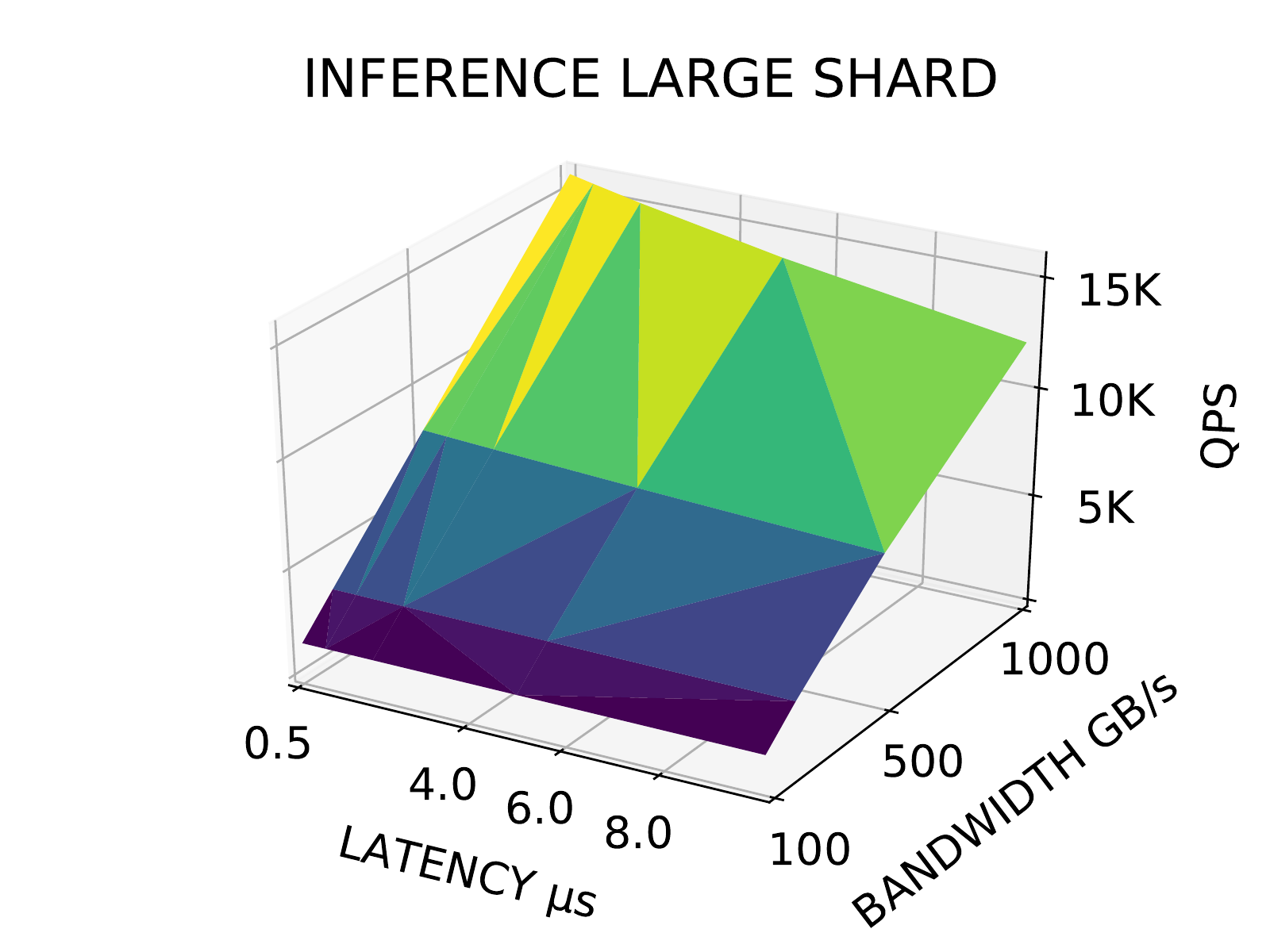} }}%
	\caption{Inference performance upper bounds as a function of Collective Communications latency and bandwidth for small/large batches, with and without Sharding. See text for analysis.}%
	\label{fig:qps_inf}%
\end{figure*}

Fig.~\ref{fig:qps_inf} shows upper bounds on achievable system throughput for
inference.
For the large batch, large embeddings case, unsharded, throughput is primarily limited by memory accesses for embeddings
such that interconnect is of secondary importance.
Such models would run well on existing systems including scale-out topologies with limited bandwidth.

\begin{figure}%
	\centering
	\subfloat[With 100GB/s Bandwidth per chip]{{\includegraphics[clip, trim=2.5cm 5cm 2.5cm 5.5cm, width=\columnwidth]{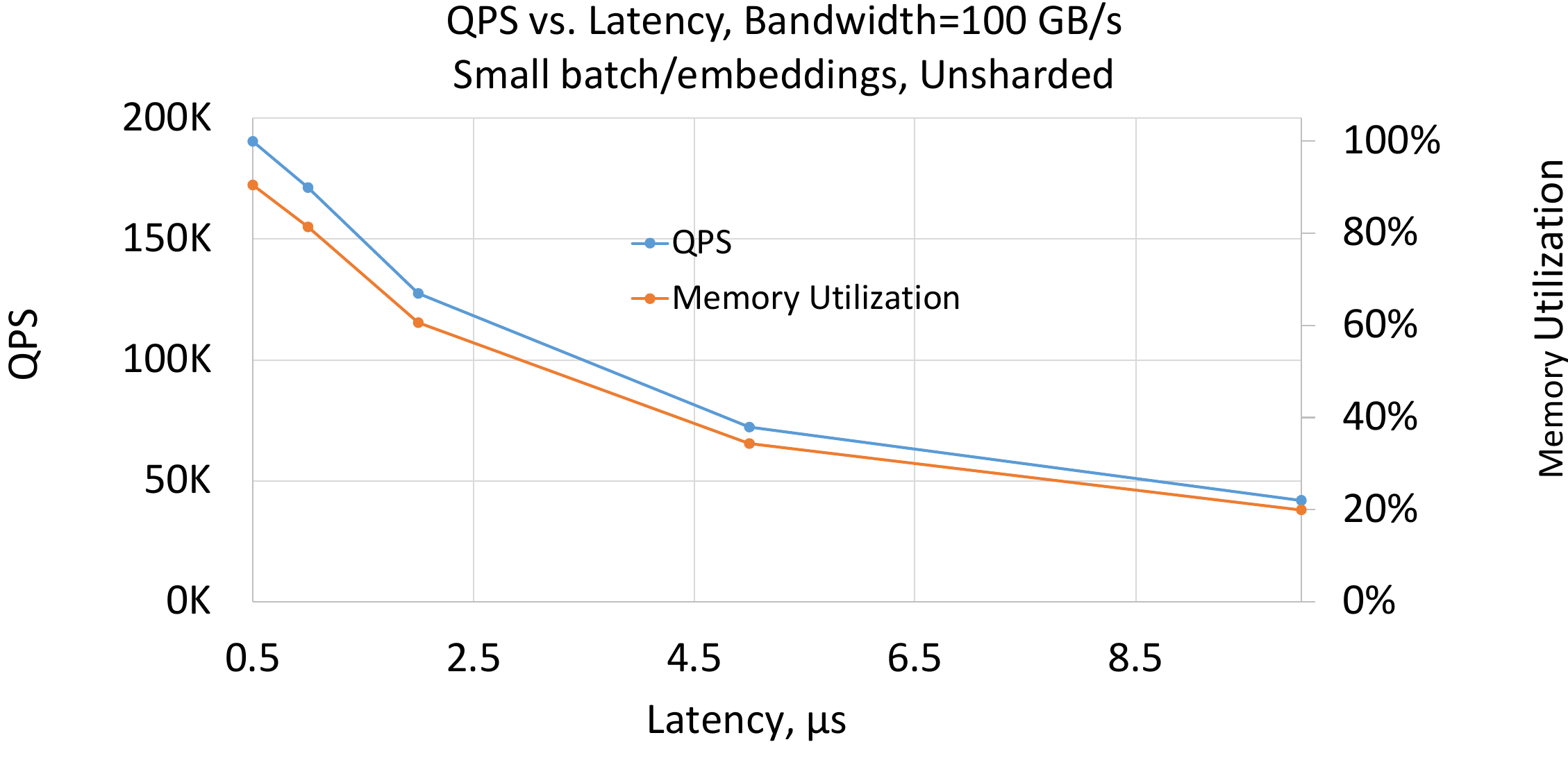} }}%
	\newline
	\subfloat[With 1,000GB/s Bandwidth per chip]{{\includegraphics[clip, trim=2.5cm 5cm 2.5cm 5.5cm, width=\columnwidth]{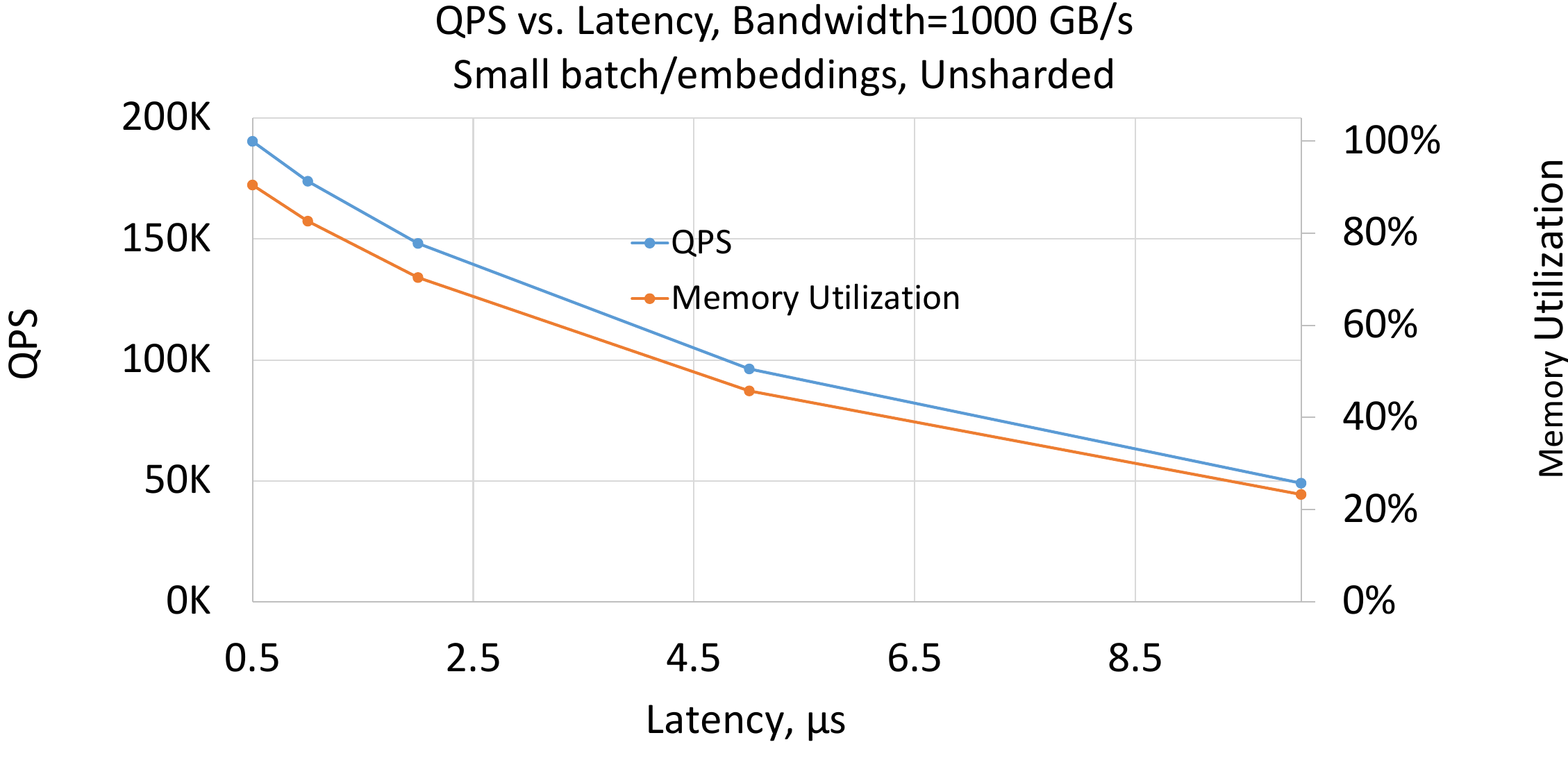} }}%
	\caption{Impact of latency on QPS for small batch size, small embedding vectors, Unsharded. Latency matters at both high and low bandwidth, and there is benefit in driving CC latency down to typical network switch port to port levels.}%
	\label{fig:insight_sm_unshard_latency}%
\end{figure}

Latency, as opposed to bandwidth, matters most for small batch/embedding workloads,
as detailed in Fig.~\ref{fig:insight_sm_unshard_latency}.
For the unsharded case this sensitivity applies at high bandwidth as well as at low bandwidth,
with throughput dropping by almost $5\times$ as latency increases. This is not surprising since
the typical all-to-all message sizes for this configuration are 320KB of indices per processor and 64KB for
pooled embeddings, small messages that would typically fall within the latency-dominated regime of CC.

This would indicate that when batch sizes are relatively small and tables are allocated to each
fit within the memory attached to each processor, it is more important to design systems to minimize latency as opposed
to pushing per-chip bandwidth. Scale-out architectures, with multiple interconnect hops and long
physical distances, make this difficult. On the other hand, ``scale-in'' system design can
help reduce latency.

The sharded, small batch/embeddings case is sensitive to both latency and bandwidth
since the exchange of unpooled embeddings results in an all-to-all message size of ${\sim}$5.2MB per processor.
However, even in this situation, there is limited benefit in improving bandwidth unless such improvement is
also accompanied by reduced latency.

The unsharded, large batch/embedding case is very slightly sensitive to latency and largely insensitive
to bandwidth. This is because the communication volume, compared to the unsharded small batch/embeddings case,
increases by 3${\times}$ for indices and by 12${\times}$ for embeddings; the resulting message sizes are still
typically within the latency-dominated region of CC. However, the number of bytes of memory lookups for embedding tables
increases by 12${\times}$ such that memory lookup time becomes the dominant term.

The sharded, large batch/embeddings case depends on both latency and bandwidth due to the larger
message sizes from exchanging unpooled embeddings.

\begin{figure}[htbp]
	\captionsetup{width=.8\linewidth}
	\centerline{\includegraphics[clip, trim=2cm 3.5cm 2.5cm 3.5cm, width=\columnwidth]{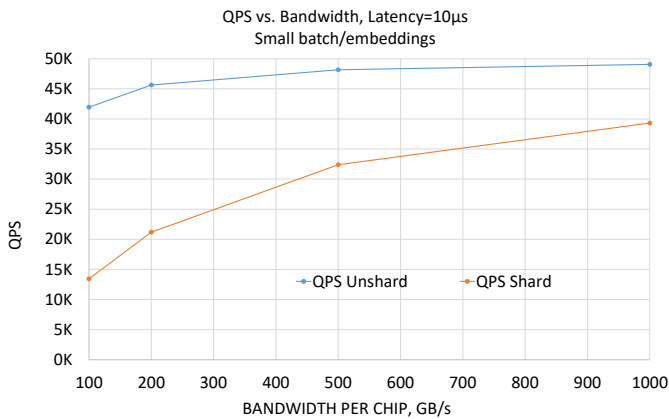}}
	\caption{High Bandwidth helps minimize the performance loss from sharding, even at high latency, due to the all-to-all exchange of unpooled embedding entries between processors.}
	\label{fig:insight_sm_bw_minimize_sharding_impact}
\end{figure}

In all cases, higher bandwidth minimizes the throughput impact of sharding.
As shown in Fig.~\ref{fig:insight_sm_bw_minimize_sharding_impact},
increasing bandwidth from 100GB/s to 1000GB/s reduces the impact of full sharding from about $3.1\times$ loss down to $1.2\times$
for the small batch/embeddings case. Similarly significant gains are seen for the large batch/embeddings case with sharding.

\subsection{Training}
\label{eval:homotrain}

\begin{figure*}%
	\centering
	\subfloat[Small batch, Small embeddings, Unsharded]{{\includegraphics[clip, trim=3.5cm 0cm 0cm 0cm, width=0.8\columnwidth]{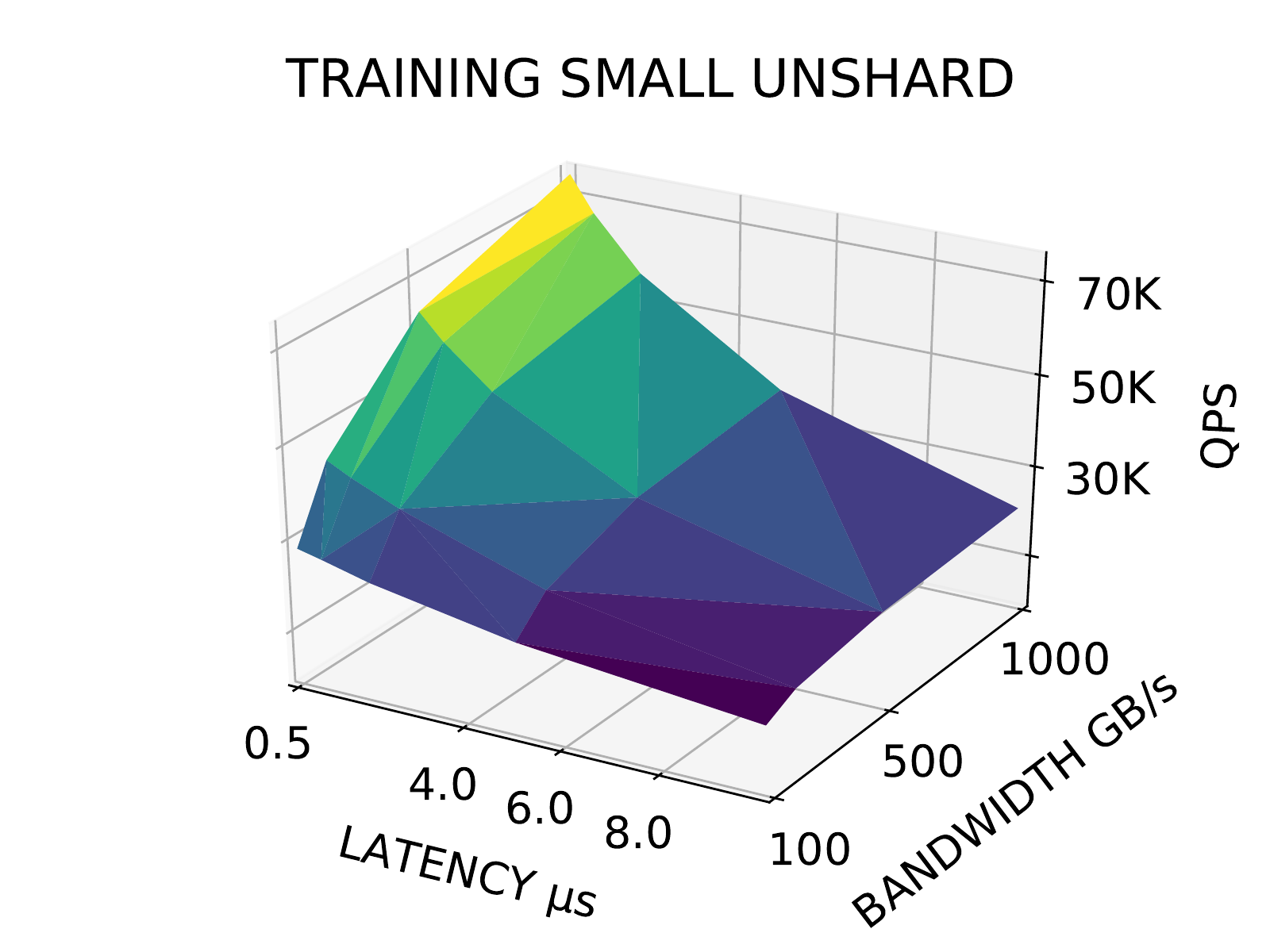} }}%
	\qquad
	\subfloat[Small batch, Small embeddings, Sharded]{{\includegraphics[clip, trim=3.5cm 0cm 0cm 0cm, width=0.8\columnwidth]{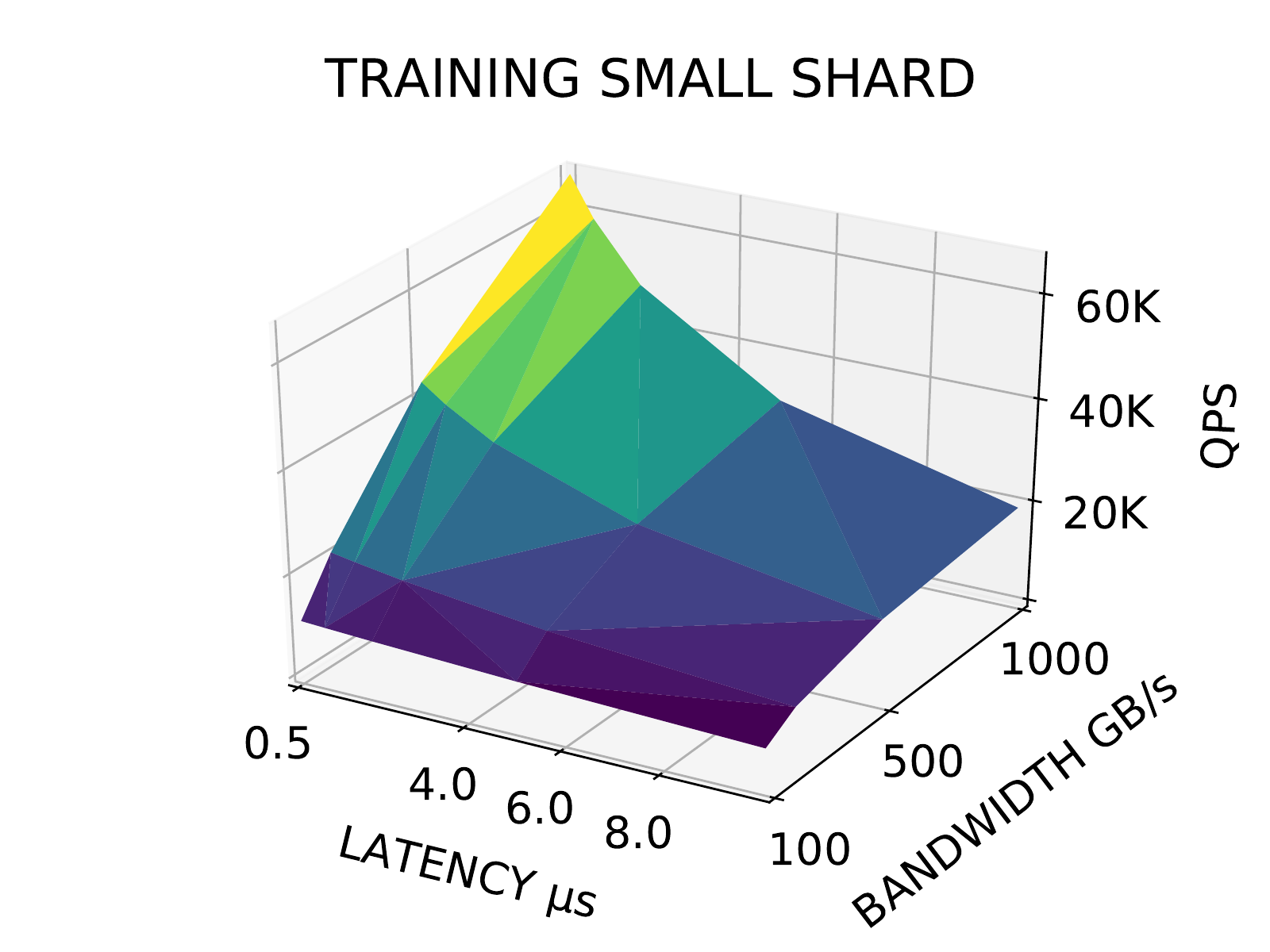} }}%
	\newline
	\subfloat[Large batch, Large embeddings, Unsharded]{{\includegraphics[clip, trim=3.5cm 0cm 0cm 0cm, width=0.8\columnwidth]{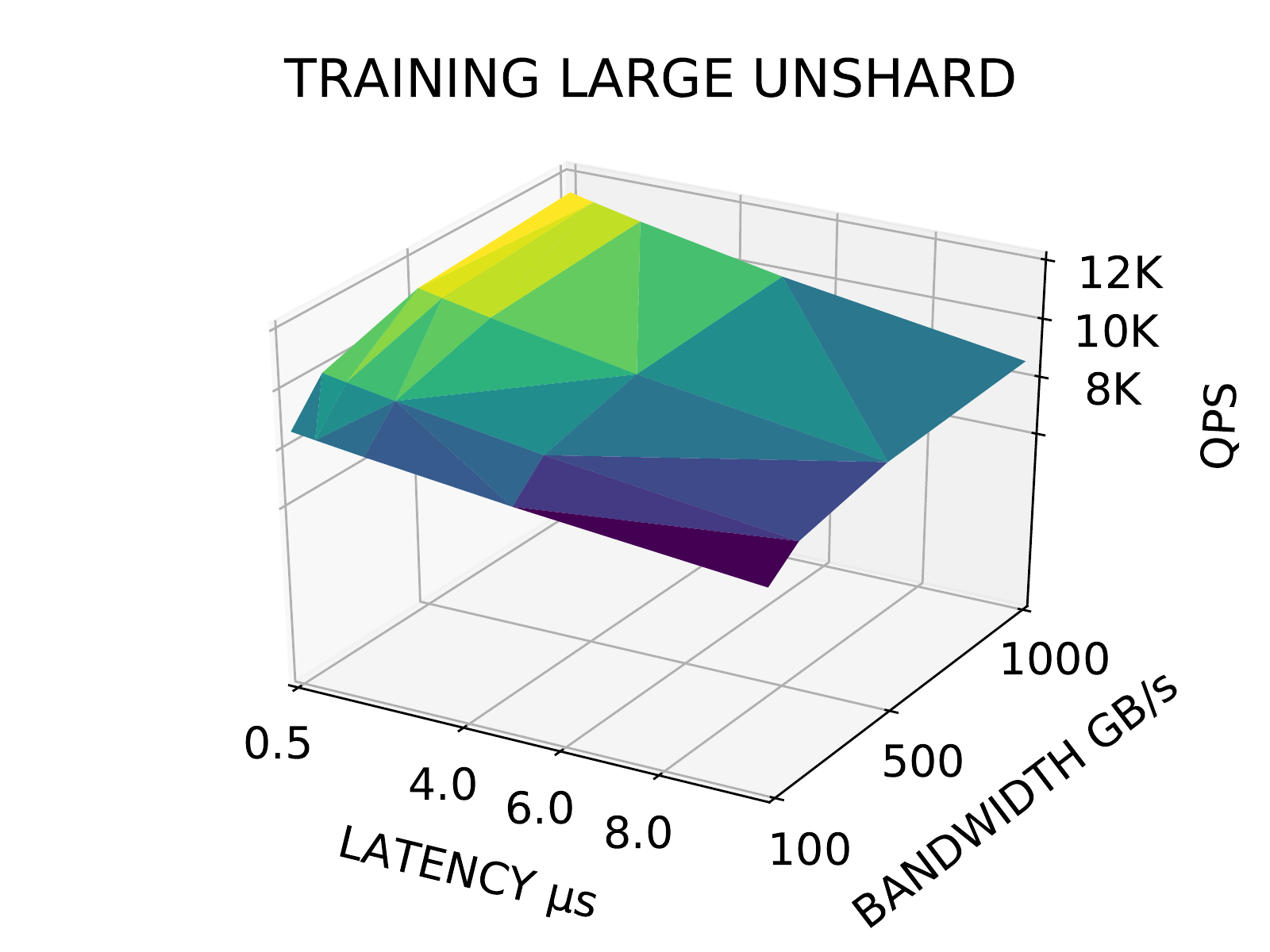} }}%
	\qquad
	\subfloat[Large batch, Large embeddings, Sharded]{{\includegraphics[clip, trim=3.5cm 0cm 0cm 0cm, width=0.8\columnwidth]{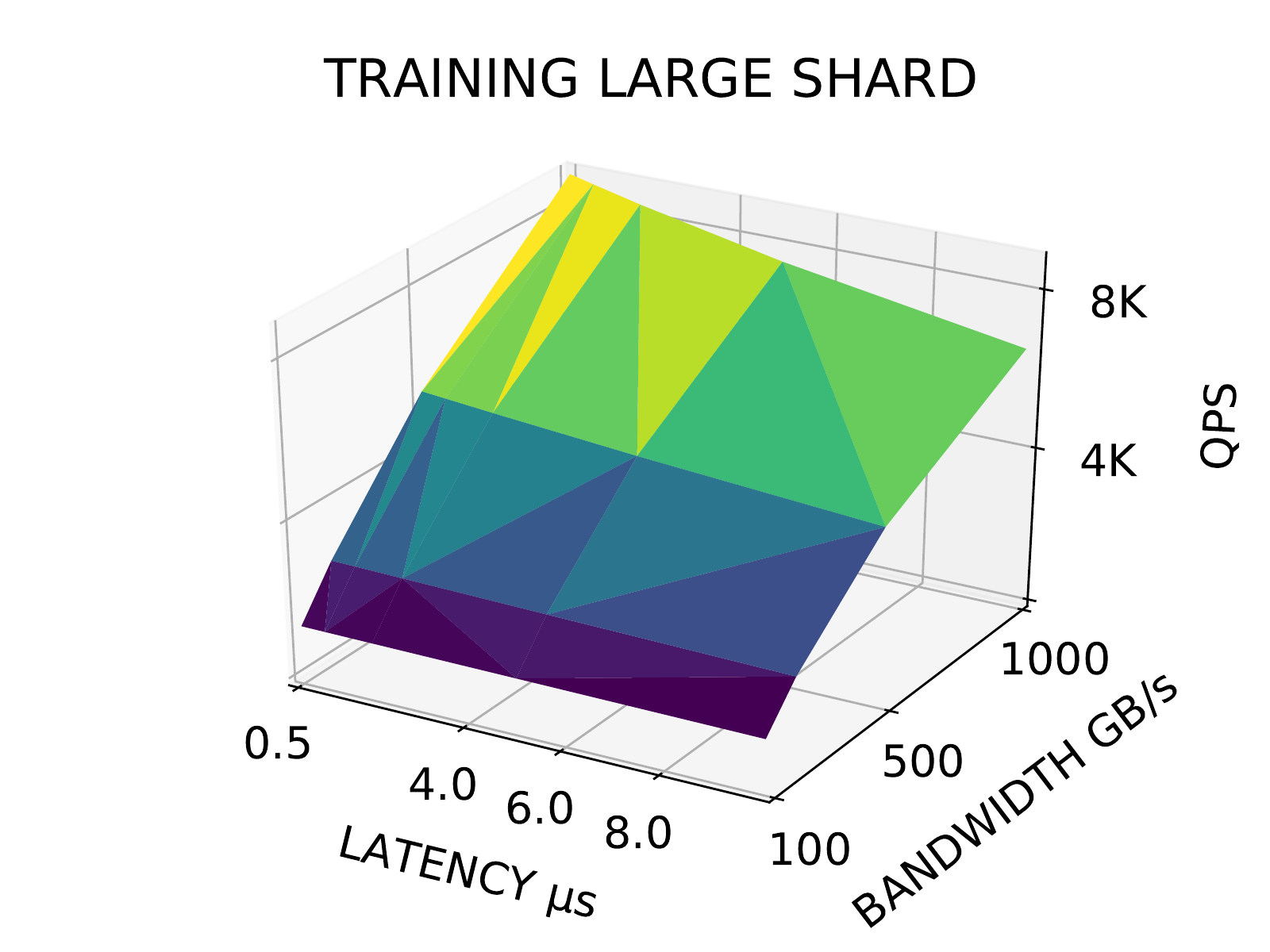} }}%
	\caption{Training performance upper bounds as a function of Collective Communications latency and bandwidth for small/large batches, with and without Sharding. See text for analysis.}%
	\label{fig:qps_train}%
\end{figure*}

\begin{figure}%
	\centering
	\subfloat[QPS]{{\includegraphics[clip, trim=2cm 4.5cm 2.5cm 5cm, width=0.9\columnwidth]{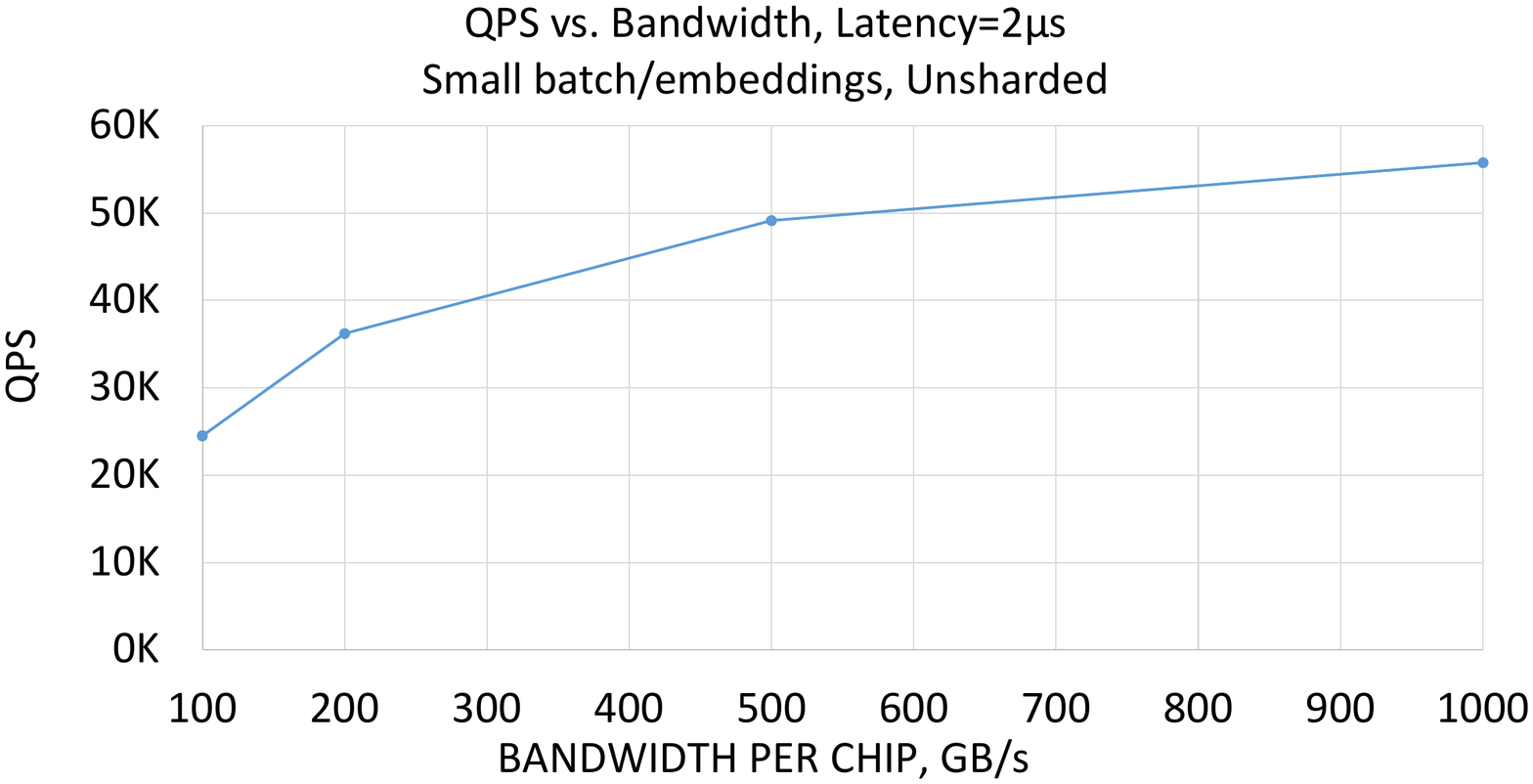} }}%
	\newline
	\subfloat[Relative duration of training phases]{{\includegraphics[clip, trim=2cm 3cm 3.5cm 3cm, width=0.9\columnwidth]{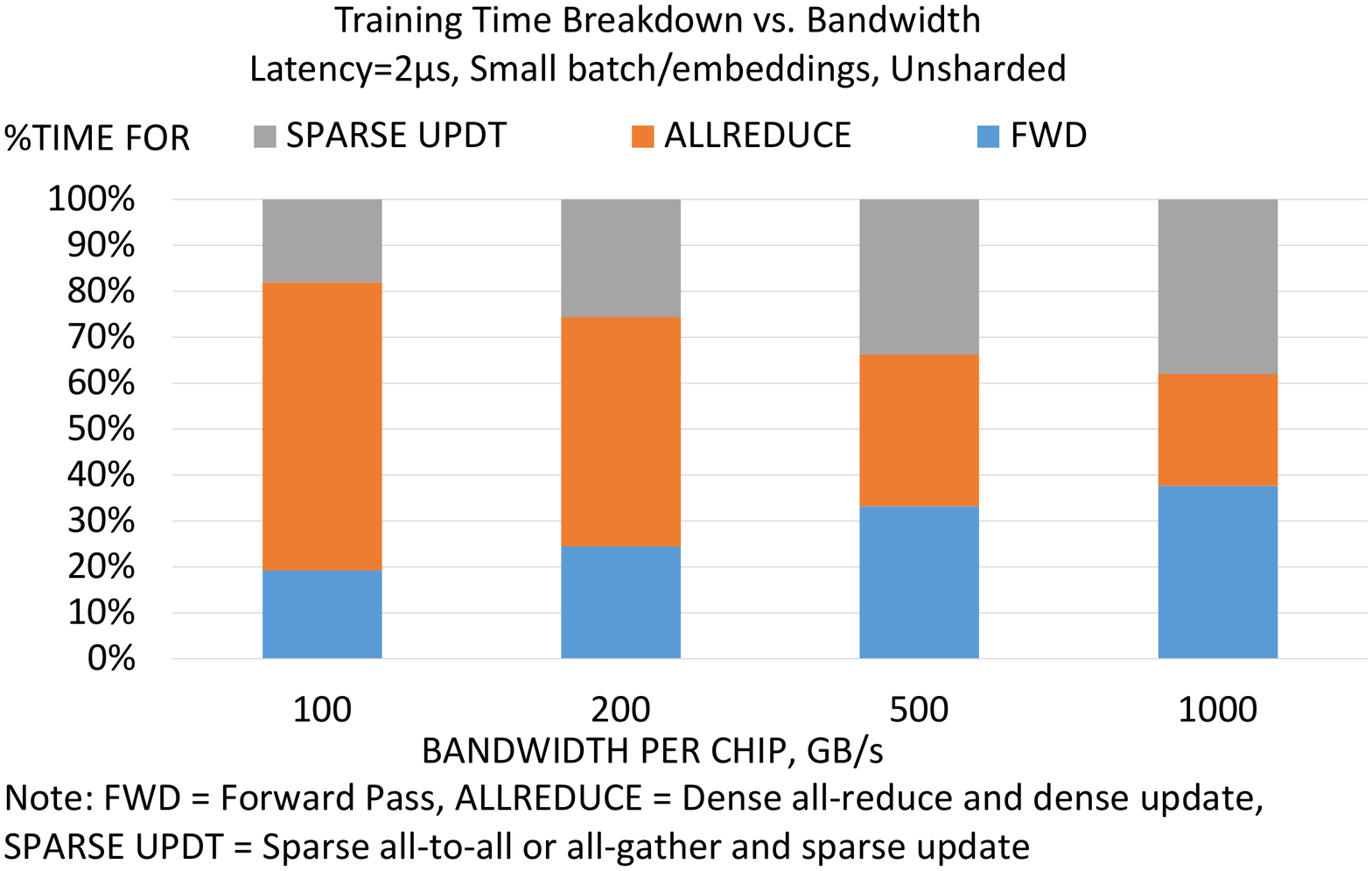} }}%
	\caption{Impact of Bandwidth on Training, small batch size, small embeddings, Unsharded. In contrast to inference, bandwidth helps directly for training due the all-reduce of large message sizes from averaging gradients across processors.}%
	\label{fig:insight_train_bwmatters_but_lat_too_hi_bw_sys_balance}%
\end{figure}

Fig.~\ref{fig:qps_train} shows upper bounds on achievable system throughput for
training. Similarly to inference, we note that the case of unsharded large batches/embeddings is
primarily memory-bound, hence does not depend much on CC latency or bandwidth.

For the other configurations, compared to inference, the importance of
optimizing bandwidth and latency are more balanced for training.
For the small batch/embeddings case, bandwidth matters to training QPS
as detailed in Fig.~\ref{fig:insight_train_bwmatters_but_lat_too_hi_bw_sys_balance}.
High bandwidth cuts dense all-reduce times since the message sizes involved are ${\sim}$2.4MB per processor;
the impact of bandwidth is, as expected, most felt at latencies under 4${\mu}$s.

\begin{figure}%
	\centering
	\subfloat[QPS]{{\includegraphics[clip, trim=2.5cm 3.5cm 2cm 3.5cm, width=0.9\columnwidth]{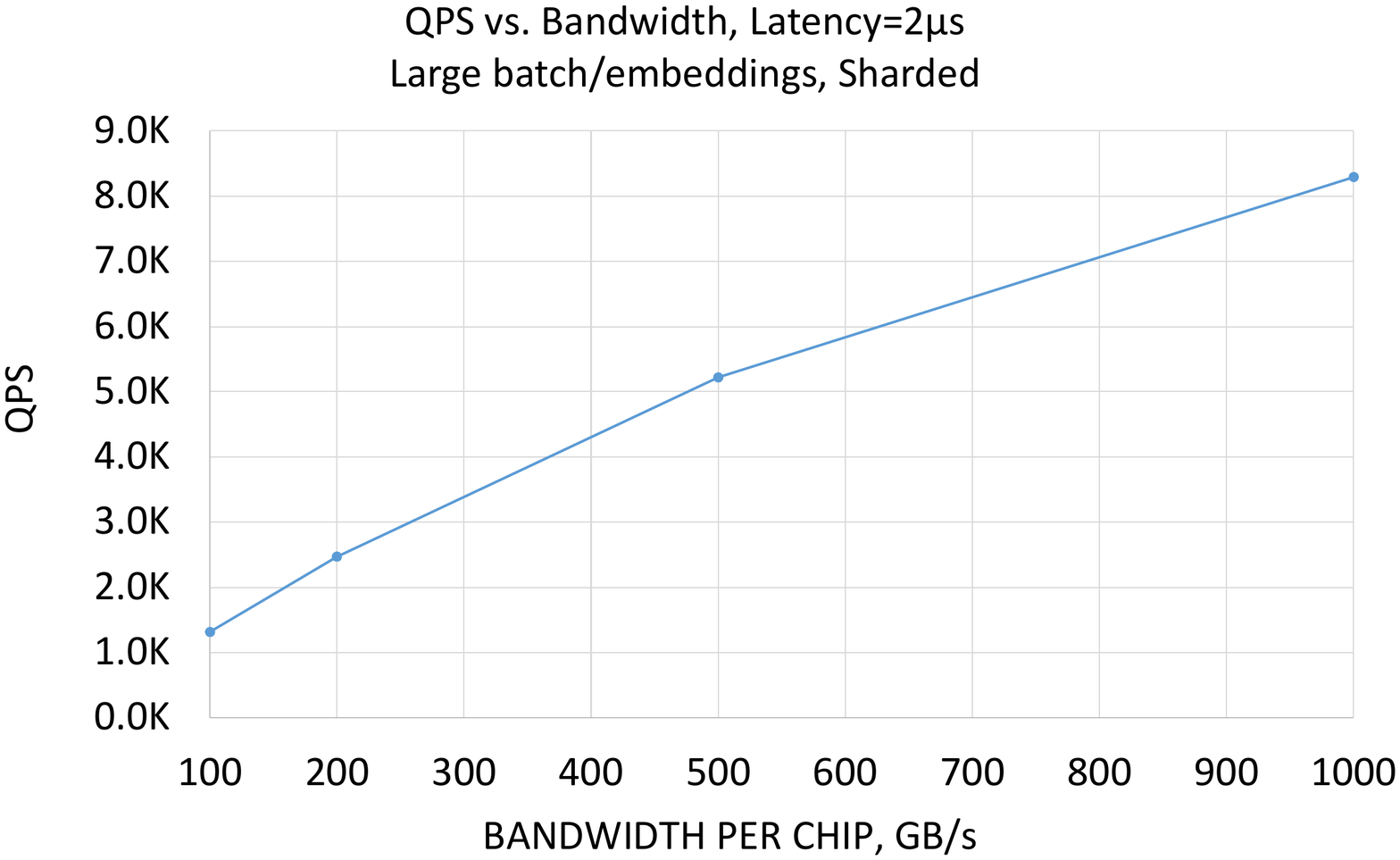} }}%
	\newline
	\subfloat[Relative duration of training phases]{{\includegraphics[clip, trim=2cm 3cm 2cm 3cm, width=0.9\columnwidth]{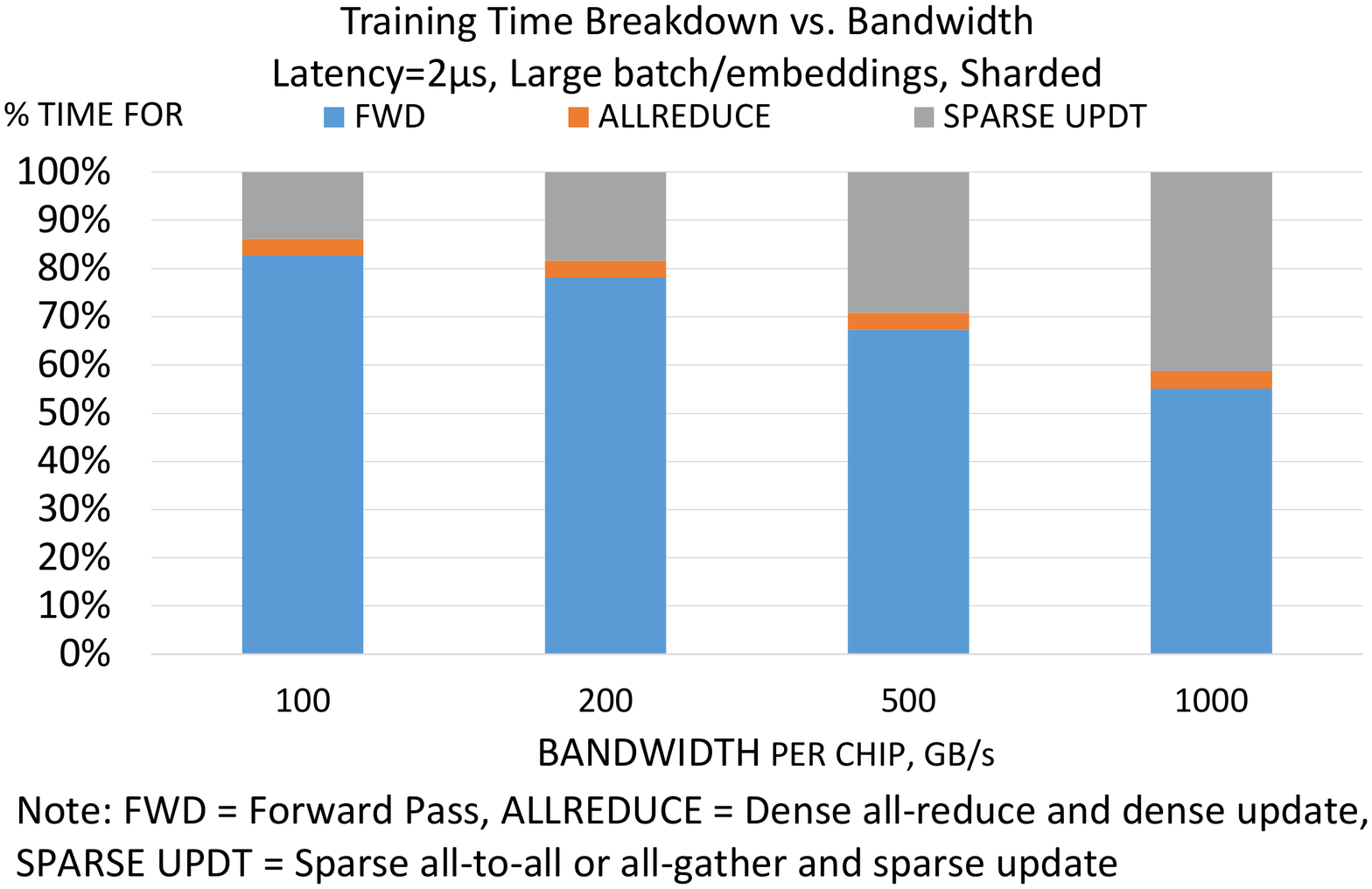} }}%
	\caption{Impact of Bandwidth on Training, large batch size, large embeddings, Sharded. In this situation, the all-to-all exchange of unpooled embeddings between processors dominates. Increasing bandwidth reduces the time spent in this all-to-all, thus speeding up the forward pass.}%
	\label{fig:insight_train_shard_bw_helps}%
\end{figure}

As with inference, higher bandwidth helps mitigate the throughput penalty
of sharding. In the large batch/embeddings scenario, overall throughput increases almost proportionally with
bandwidth as shown in Fig.~\ref{fig:insight_train_shard_bw_helps}.
The all-to-all exchange of unpooled embeddings between processors dominates,
with message sizes of ${\sim}$60MB per processor.

\section{RecSpeed: An Optimized System Architecture for RecSys}
\label{recspeed}

This section describes the features of RecSpeed, a hypothetical system architecture for RecSys workloads. The objectives of RecSpeed are to:

\begin{itemize}
	\item Maximize throughput for inference and training of large RecSys models;
	\item Support future, ever-larger RecSys models;
	\item Allow implementation using existing process technologies,
	and to fit into common datacenter power envelopes;
	\item Support existing SW and HW interfaces for datacenter AI server racks.
\end{itemize}

\subsection{RecSpeed Architecture Features}
\label{recspeed:design}

\begin{figure}%
	\centering
	\subfloat[Schematic of chip]{{\includegraphics[clip, trim=16.5cm 10.5cm 8.5cm 4.5cm, width=0.65\columnwidth]{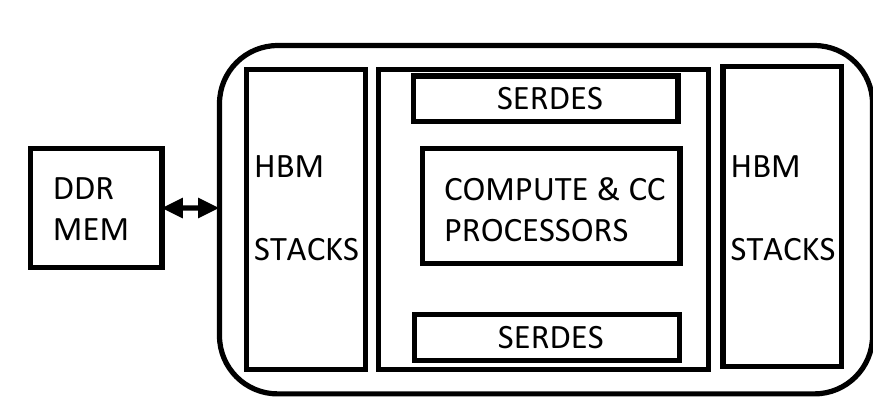} }}%
	\newline
	\subfloat[System interconnect example with 6 chips]{{\includegraphics[clip, trim=2cm 6cm 22cm 4cm, width=0.7\columnwidth]{recspeed.pdf} }}%
	\caption{Proposed RecSpeed Architecture.}%
	\label{fig:recspeed}%
\end{figure}

Fig.~\ref{fig:recspeed} shows a sketch of the proposed chip and interconnect structure for a 6-chip RecSpeed system.
Key features of the architecture are as follows:

\begin{itemize}
	\item Fixed-topology quadratic point-to-point interconnect without any form of switching to minimize latency.
	\item HW support for Collective Communications to minimize synchronization and SW-induced latency.
	\item Fast HBM memory attached to each chip, as many stacks as practical;
	as of the writing of this paper, NVIDIA's A100 has room for 6 stacks of HBM (of which only 5 are used),
	each 4-deep; however, 8-deep stacks
	are available.
	\item Slow bulk memory, such as DDR4.
	\item Optimized packaging and system design to allow ``scale-in'', packing as many RecSpeed
	chips as possible in close physical proximity.
\end{itemize}

High-density physical packaging of chips is particularly important in order to achieve maximum throughput;
when interconnect moves
from intra-board to the system level, the energy consumed per bit goes up by $25\times$,
bandwidth falls by over $20\times$, and overhead increases markedly \cite{mcmgpu}.

\textbf{Fixed-topology vs. switched all-to-all interconnect:} The proposed interconnect for RecSpeed
could reduce latency compared to the all-to-all switched interconnect found in
NVIDIA's DGX-2 \cite{nvswitch} or Habana's HLS-1 training system \cite{habanahc}.
Specifically, the presence of a switch introduces several hundred nanoseconds of
additional latency \cite{teralynx}\cite{intelsw}.
A quadratic interconnect offers performance gains of $2.3{\times}$ for large CC all-to-all
messages on an 8-node system, compared to a ring interconnect,
and for smaller message sizes the gain can range as high as $15{\times}$ \cite{fbscaleupscaleout}.

\textbf{HW support for CC:} We note that the proposed interconnect structure can
efficiently support both CC all-to-all and all-reduce operations with minimal latency
and bandwidth usage that matches the theoretical lower bound \cite{ccbook}.

\textbf{Implementing High-Bandwidth Links:} The proposed high-bandwidth RecSpeed
links can be implemented via existing technology, amounting to ${\sim}31\%$ of
the bandwidth of a Tomahawk 4 switch chip \cite{th4}.

\textbf{Hybrid Memory Support:} Certain embedding tables and vectors are accessed
less often than others \cite{aibox}. It is therefore useful to provide a two-level
memory system, with large bulk memory in the form of DDR4 combined with fast HBM memory.
Tables can be allocated to one memory or the other statically{\textemdash}sharded or not{\textemdash}or the faster
HBM can be used as a cache. Static allocation is preferred, as dealing with such
a large cache structure where the smaller memory has latency comparable to that of the larger memory
may not offer much benefit, based on prior efforts such as Intel's
Knights Landing HPC architecture \cite{knlhbm}. With the configuration shown,
up to 5.5TB of memory can be provided for a RecSpeed system, of which 27\% would
be fast HBM. Baidu reports that their AIBox \cite{aibox} system is able to effectively hide storage
access time, despite the two order of magnitude latency difference and the greater than one order
of magnitude bandwidth difference between SSD and main memory. Since the performance gap
between DDR and HBM memory is smaller, it is reasonable to assume that careful system design can
enable a hybrid memory system to run closer to the performance of a full HBM system.

\subsection{RecSpeed Performance vs. DGX-2}
\label{recspeed:perf}

\begin{table}[htbp]
	\caption{Numbers used for RecSpeed performance upper bounds.}
	
	\begin{center}
		\begin{tabular}{L{0.25\columnwidth} L{0.6\columnwidth}}
			\hline
			\textbf{Parameter} & \textbf{Value}\\
			\hline \hline
			CC all-to-all & Latency of 1000ns, bandwidth of 1000GB/s\\
			\hline
			CC all-reduce & Latency of 1000ns, bandwidth of 1000GB/s\\
			\hline
			HBM Memory & HBM2E @ 3000MHz, 6 stacks, 96GB\\
			\hline
			DDR4 Memory & 1-channel, up to 256GB 3D TSV DIMM \\
			\hline
			\#chips/system & 16\\
			\hline
			Compute capability & 200 TFLOPS FP16\\
			\hline
			Memory Size & 1.5TB HBM2E + 4TB DDR4\\
			\hline
		\end{tabular}
	\end{center}
	\label{table:rsdata}
\end{table}

\begin{table}[htbp]
	\caption{Numbers used for NVIDIA DGX-2 performance upper bounds.}
	
	\begin{center}
		\begin{tabular}{L{0.4\columnwidth} L{0.45\columnwidth}}
			\hline
			\textbf{Parameter} & \textbf{Value}\\
			\hline \hline
			CC all-to-all & Derived from CC all-gather \cite{dgxlat}\\
			\hline
			CC all-reduce, CC all-gather & Data from \cite{dgxlat}\\
			\hline
			Memory System & HBM2, 4 stacks @ 2300MHz\\
			\hline
			Bandwidth per chip & 150GB/s\\
			\hline
			\#chips/system & 16\\
			\hline
			Compute capability & 125 TFLOPS FP16\\
			\hline
			\emph{On-chip memory} & \emph{Assumed sufficient{\textemdash}not the case in practice}\\
			\hline
		\end{tabular}
	\end{center}
	\label{table:dgx2data}
\end{table}

Table~\ref{table:rsdata} shows the system characteristics that we use for computing the performance upper bound for RecSpeed, and Table~\ref{table:dgx2data} for the DGX-2. Note that we assume a ``modified'' V100 chip with
more on-chip buffering memory than the actual V100.

The resulting throughput numbers and comparison versus NVIDIA's DGX-2 estimated upper bounds
are shown in Table~\ref{table:rsinf} for inference and Table~\ref{table:rstrain} for training.
In our model, the DGX-2 is largely bound by its high CC latency, which can likely be reduced via software optimization.

\begin{table}[htbp]
	\caption{RecSpeed Inference Upper Bounds.}
	
	\begin{center}
		\begin{tabular}{lccccc}
			\hline
			Config. & QPS & Mem. & DGX-2 & DGX-2 & Potential\\
			& & Util. & QPS & Mem.Util. & Speedup\\
			\hline \hline
			Sm. Unshard & 300K & 67\% & 4.9K & 1.8\% & $62\times$\\
			\hline
			Sm. Shard & 207K & 47\% & 4.5K & 1.6\% & $46\times$\\
			\hline
			Lg. Unshard & 56K & 93\% & 4.7K & 15\% & $12\times$\\
			\hline
			Lg. Shard & 30K & 50\% & 2.1K & 7\% & $14\times$\\
			\hline
		\end{tabular}
	\end{center}
	\label{table:rsinf}
\end{table}

\begin{table}[htbp]
	\caption{RecSpeed Training Upper Bounds.}
	
	\begin{center}
		\begin{tabular}{lccccc}
			\hline
			Config. & QPS & Allred. & DGX-2 & DGX-2 & Potential\\
			& & \%Time & QPS & Allred. & Speedup\\
	\hline \hline
	Sm. Unshard & 99K & 33\% & 2.2K & 31\% & $45\times$\\
	\hline
	Sm. Shard & 83K & 28\% & 2.1K & 30\% & $39\times$\\
	\hline
	Lg. Unshard & 25K & 9\% & 2K & 28\% & $12\times$\\
	\hline
	Lg. Shard & 16K & 6\% & 1.2K & 18\% & $13\times$\\
	\hline
	\multicolumn{6}{p{.8\columnwidth}}{All-reduce refers to dense reduction and update.}\\
\end{tabular}
\end{center}
\label{table:rstrain}
\end{table}

\textbf{Limitations: }In this section, we do not discuss trade-offs and issues{\textemdash}important as they
may be{\textemdash}relating to die size, power, and thermal design.

\section*{Conclusion}
This paper reviews the features of a representative Deep Learning Recommender System
and describes hardware architectures that are used to deploy this and similar
workloads. The performance of this representative Deep Learning Recommender is
investigated with respect to its sensitivity to hardware system design
parameters for training and inference.

We identify the \emph{latency} of collective communications operations as
a crucial, yet overlooked, bottleneck that can limit recommender system throughput
on many platforms.
We also identify per-chip communication bandwidth,
on-chip buffering
and memory system lookup rates as further factors.

We show that a novel architecture could achieve substantial throughput gains
for inference and training on recommender system workloads by improving these factors beyond state of the art
via the use of ``scale-in'' to pack processing chips in close physical proximity,
a two-level high-performance memory system
combining HBM2E and DDR4, and a quadratic point-to-point fixed topology interconnect.
Specifically, achieving CC latencies of 1$\mu$s and chip-to-chip bandwidth of 1000GB/s
would offer the potential to boost recommender system
throughput by $\boldsymbol{12}$--$\boldsymbol{62\times}$ for inference and
$\boldsymbol{12}$--$\boldsymbol{45\times}$ for training
compared to upper bounds for NVIDIA's DGX-2 large-scale AI platform,
while minimizing the performance impact of ``sharding'' embedding tables.

\section*{Acknowledgment}

The authors would like to thank Prof. Kurt Keutzer and Dr. Amir Gholami for their
support and their constructive feedback.

\newpage
\ 
\newpage
\appendix[Steps and Operators for DLRM]
\label{appx}

This appendix presents a high-level algorithm representing the steps for training a DLRM model in a distributed fashion with $n$ identical processors. The forward pass is covered in algorithm \ref{alg:dlrm_fwd_steps} (used in both inference and training, note that for training it is assumed that activations are checkpointed as needed and optimally \cite{checkmate} to facilitate backpropagation) and the backward pass and weight update in algorithm \ref{alg:dlrm_bckwd_steps}. Note that most operations are performed concurrently across all processors as the inference query or training batch is split up across all $n$ processors. For simplicity, it is also assumed that the concatenation of the bottom MLP output and the pairwise dot products (after duplicate removal, diagonal removal, and vectorizing is completed) is performed as part of the \textbf{FeatureInteractions} operation.

For training, the optimizer used is vanilla SGD.
AdaGrad is reported to achieve slightly better results for DLRM on the Criteo Ad Kaggle
dataset \cite{fbdlrm}, however we use vanilla SGD in our steps and performance model for consistency with the DLRM repo.
The pipelining during the dense backward pass of collective communications (i.e. all-reduce operations) with the backpropagation computations and parameter updates is not shown; in practice, this is certainly feasible as long as the all-reduce latency is acceptable. 

The notations used are shown in Table~\ref{table:intermediate_vars}.

\begin{table}[htbp]
	\caption{Notations for DLRM steps.}
	
	\begin{center}
		\begin{tabular}{L{0.15\columnwidth} L{0.15\columnwidth} L{0.55\columnwidth}}
			\hline
			\textbf{Symbol} & \textbf{Usage} & \textbf{Meaning}\\
			\hline \hline
			$n_d$ & Constant & Number of dense features in the model\\
			\hline
			$n_c$ & Constant & Number of sparse features or embedding tables\\
			\hline
			$c_i$ & Constant & Cardinality of the $i$th categorical feature\\
			\hline
			$l_i$ & Constant & Number of lookups performed on the $i$th embedding table\\
			\hline
			$l_b$ & Constant & Number of bottom MLP layers\\
			\hline
			$l_t$ & Constant & Number of top MLP layers\\
			\hline
			$d$ & Constant & Embedding dimension\\
			\hline
			$O_B$ & Fwd & Output of bottom MLP up to a given layer\\
			\hline
			$L_i$ & Fwd & Local embedding lookup indices for $i$th table after indices all-to-all\\
			\hline
			$O_E$ & Fwd & Local embedding lookup vectors (possible pooled)\\
			\hline
			$V_i$ & Fwd & Pooled embedding vectors (after second all-to-all) resulting from lookups for the $i$th table\\
			\hline
			$F$ & Fwd & Feature interactions input denoted as $A$ in feature interactions layer description of Sec.~\ref{RecSysOverview:Models}\\
			\hline
			$O_T$ & Fwd & Output of bottom MLP up to a given layer\\
			\hline
			${GT}_i$ & Bckwd/update & Gradient of loss w.r.t. output of $i$th top MLP FC layer\\
			\hline
			${GB}_i$ & Bckwd/update & Gradient of loss w.r.t. output of $i$th bottom MLP FC layer\\
			\hline
			${GE}_i$ & Bckwd/update & Pooled gradient (i.e. on processor which doesn't own relevant embeddings) of loss w.r.t. lookups from $i$th embedding table\\
			\hline
			${LGE}_i$ & Bckwd/update & Pooled gradient of loss w.r.t. lookups from $i$ table on processor which owns relevant embeddings after all-to-all/all-gather\\
			\hline
			${FGE}_i$ & Bckwd/update & Unpooled/expanded batch-reduced gradient of loss w.r.t. lookups from $i$ table on processor which owns relevant embeddings after all-to-all/all-gather\\
			\hline
		\end{tabular}
	\end{center}
	\label{table:intermediate_vars}
\end{table}

\begin{algorithm}
	\SetAlgoLined
	
	\KwIn{Number of processors $p$; $b{\times}n_d$ batch of dense features $D$; $n_c$ sets of sparse features each one called $S_i$, each one $b{\times}l_i$; bottom MLP layers ${FC}_{B_i}$, $i \in [1, l_b]$; top MLP layers ${FC}_{T_i}$, $i \in [1, l_t]$; $n_c$ embedding tables denoted as $E_i$ with each table representing a $c_i{\times}d$ matrix.}
	
	$O_B \leftarrow D$\\
	
	\For{$i = 1 \cdots l_b$}{
		$O_B \leftarrow {FC}_{B_i}(O_B)$
	}
	
	$L_1, L_2, \cdots, L_{n_c} \leftarrow {all\_to\_all}([S_1, S_2, \cdots, S_{n_c}])$
	
	$O_E \leftarrow []$\\
	\For{$j = 1 \cdots n_c$}{
		$O_{E_j} = E_j(L_j)$\\
		\If{no\_sharding}{$O_{E_j} = pool(O_{E_j})$}
		${O_E}.append(O_{E_j})$
	}
	
	\If{full\_sharding}{
		$V_1, V_2, \cdots, V_{n_c} \leftarrow reduce\_scatter(O_E)$\\
	}
	\If{no\_sharding}{
		$V_1, V_2, \cdots, V_{n_c} \leftarrow all\_to\_all(O_E)$
	}
	
	$F \leftarrow concat([O_B, V_1, V_2, \cdots, V_{n_c}])$\\
	
	$F' \leftarrow FeatureInteractions(F)$
	
	$O_T \leftarrow F'$\\
	
	\For{$i = 1 \cdots l_t$}{
		$O_T \leftarrow {FC}_{T_i}(O_T)$
	}
	
	$\mathbf{p} \leftarrow O_T[:, 0]$
	
	\KwOut{Predicted click probabilities vector $\mathbf{p}$}
	
	\caption{DLRM forward pass steps.}
	\label{alg:dlrm_fwd_steps}
\end{algorithm}

\begin{algorithm}
	\SetAlgoLined
	
	\KwIn{Number of processors $p$; $b{\times}n_d$ batch of dense features $D$; $n_c$ sets of sparse features each one called $S_i$, each one $b{\times}l_i$; bottom MLP backward operators ${FCBackward}_{B_i}$, $i \in [1, l_b]$; top MLP backward operators ${FCBackward}_{T_i}$, $i \in [1, l_t]$; $n_c$ embedding tables denoted as $E_i$ with each table representing a $c_i{\times}d$ matrix. Learning rate $\gamma$; predictions $\mathbf{p} \in (0, 1)^n$; labels $\mathbf{l} \in [0, 1]^n$; loss function $\mathcal{L}(\mathbf{p}, \mathbf{l}) = \frac{1}{n} \sum_{i = 1}^{n} l_i \log p_i + (1 - l_i) \log (1 - p_i)$ with predictions $\mathbf{p}$ and labels $\mathbf{l}$.}
	
	$L_{BCE} \leftarrow \mathcal{L}(\mathbf{p}, \mathbf{l})$\\
	
	$\nabla_{\textbf{p}} {\mathcal{L}} \leftarrow \frac{1}{n} \sum_{i = 1}^{b} (\frac{l_i}{p_i} - \frac{1 - l_i}{1 - p_i})$\\
	
	${GT}_{l_t + 1} \leftarrow \nabla_{\textbf{p}} {\mathcal{L}}$\\
	\For{$i = l_t, l_t - 1, \cdots, 1$}{
		${GT}_{i}, \nabla_{FC_{T_i}} \mathcal{L} \leftarrow FCBackward_{T_i}({GT}_{i + 1})$
	}
	
	${GB}_{l_b + 1}, {GE}_1, \cdots, {GE}_{n_c} \leftarrow FeatureInteractionsBackward({GT}_{1})$\\
	
	\If{no\_sharding}{
		${LGE}_1, \cdots, {LGE}_{n_c} \leftarrow all\_to\_all([{GE}_1, \cdots, {GE}_{n_c}])$\\
	}
	
	\If{full\_sharding}{
		${LGE}_1, \cdots, {LGE}_{n_c} \leftarrow all\_gather([{GE}_1, \cdots, {GE}_{n_c}])$\\
	}
	
	${FGE}_1, \cdots, {FGE}_{n_c} \leftarrow expand\_sparse\_grads([{LGE}_1, \cdots, {LGE}_{n_c}])$
	
	\For{$i = l_b, l_b - 1, \cdots, 1$}{
		${GB}_{i}, \nabla_{FC_{B_i}} \mathcal{L} \leftarrow FCBackward_{B_i}({GB}_{i + 1})$
	}
	
	${GB}_{1}, {GB}_{2}, \cdots, {GB}_{l_b}, {GT}_{1}, {GT}_{2}, \cdots, {GT}_{l_t} \leftarrow \frac{1}{p} all\_reduce([{GB}_{1}, {GB}_{2}, \cdots, {GB}_{l_b},$
	${GT}_{1}, {GT}_{2}, \cdots, {GT}_{l_t}])$
	
	\For{$i = 1 \cdots l_b$}{
		$Params(FC_{B_i}) \leftarrow Params(FC_{B_i}) - \gamma \nabla_{FC_{B_i}} \mathcal{L}$
	}
	
	\For{$i = 1 \cdots l_t$}{
		$Params(FC_{T_i}) \leftarrow Params(FC_{T_i}) - \gamma \nabla_{FC_{T_i}} \mathcal{L}$
	}
	
	\For{$i = 1 \cdots n_c$}{
		$Params(E_i)\leftarrow Params(E_i) - \gamma {FGE}_{i}$
	}
	
	\KwOut{Current loss $L_{BCE}$}
	
	\caption{DLRM backward pass and weight update steps.}
	\label{alg:dlrm_bckwd_steps}
\end{algorithm}

\textbf{Other operations}

This section introduces additional operations (other than collective communications operations) mentioned in algorithms \ref{alg:dlrm_fwd_steps} and \ref{alg:dlrm_bckwd_steps}.

\textbf{FC}: The forward pass of the FC layer.

\textbf{FeatureInteractions}: The forward pass of the dot-product feature interactions layer with exclusion of the diagonal feature interactions matrix entries.

\textbf{Concat}: The concatenation operation of the batched bottom MLP output and batched pooled embedding vectors along a new dimension as mentioned in Sec.~\ref{RecSysOverview:Models}.

\textbf{FCBackward}: This operator takes in the gradient of the loss with respect to the output of a given FC layer, uses the checkpointed input activations to the layer, and returns both the gradient of the loss with respect to the weights of the layer as well as the gradient of the loss with respect to the input to the layer. This will then be used by the next FCBackward operator.

\textbf{FeatureInteractionsBackward}: This operator takes in the gradient of the loss with respect to the output of the feature interactions layer and returns the gradient of the loss with respect to each of the batched inputs to the feature interactions layer, which are the output of the bottom MLP as well as the pooled embeddings resulting from the embedding lookups in the model. Note that there are no weights in the feature interactions layer so this is sufficient for FeatureInteractionsBackward.

\textbf{Expand\_sparse\_grads}: Because of pooling in this model, all of the gradients on embeddings that are pooled into a single output are identical. After communicating only the gradients on the pooled vectors, these values are simply ``expanded'', or copied, to every unpooled vector that was summed to generate the pooled vector. This operation also averages the gradients on the pooled vectors across the batch.

\textbf{Params}: Shorthand operator to denote the parameters associated with a given FC layer or embedding table.

\end{document}